%% file: main.tex
\documentclass[conference]{IEEEtran}
\IEEEoverridecommandlockouts
\usepackage{amsthm}
\usepackage{amsfonts,amsmath}
\usepackage{algorithm}
\usepackage[noend]{algpseudocode}
\usepackage{bm}
\usepackage{booktabs}
\usepackage{soul}
\usepackage{multirow}
\usepackage{subfig}
\usepackage[flushleft]{threeparttable}
\usepackage{adjustbox}
\usepackage{fancyhdr}
\usepackage[dvipsnames]{xcolor}
\usepackage[flushleft]{threeparttable}
\usepackage{tikz}
\usepackage{outlines}
\usepackage{balance}
\usepackage{tcolorbox}
\usepackage{wrapfig}
\usepackage{blindtext}
\usepackage{xcolor}
\usepackage{balance}
\usepackage{algorithm}
\usepackage{color}
\usepackage{listings}
\usepackage{color}
\usepackage{MnSymbol,wasysym}
\usepackage{marvosym}
\usepackage[switch]{lineno}
\usepackage{xcolor,colortbl}

\newcommand*\circled[1]{\tikz[baseline=(char.base)]{
            \node[shape=circle,fill,inner sep=0.8pt] (char) {\textcolor{white}{#1}};}}

\def\BibTeX{{\rm B\kern-.05em{\sc i\kern-.025em b}\kern-.08em
    T\kern-.1667em\lower.7ex\hbox{E}\kern-.125emX}}

\definecolor{Gray}{gray}{0.85}
\definecolor{LightCyan}{rgb}{0.88,1,1}

\usepackage{algorithm}
\usepackage{algpseudocode} 

\usepackage{twoopt}
\usepackage{tikz}
\usetikzlibrary{fit}

\newcommand\tikzmark[1]{%
  \tikz[remember picture,overlay]\node[inner xsep=0pt] (#1) {};}
\newcommandtwoopt\Textbox[5][2.5cm][2cm]{%
\begin{tikzpicture}[remember picture,overlay]
  \coordinate (aux) at ([xshift=#1]#4);
  \node[inner ysep=5pt,yshift=0ex,draw=black,
    fit=(#3) (aux),baseline] 
    (box) {};
  \node[text width=#2,anchor=north east,
    font=\sffamily\footnotesize,align=right] 
    at (box.north east) {#5};
\end{tikzpicture}%
}
\DeclareCaptionSubType*{algorithm}

\DeclareCaptionLabelFormat{alglabel}{Alg.~#2}
    
\begin{document}

\title{Dynasparse: Accelerating GNN Inference through \underline{Dyna}mic \underline{Spars}ity \underline{E}xploitation
}

\author{
\IEEEauthorblockN{ Bingyi Zhang, Viktor Prasanna}
\IEEEauthorblockA{
    University of Southern California, 
    Los Angeles, California, USA\\
    bingyizh@usc.edu, prasanna@usc.edu }
}

\maketitle

\begin{abstract}
Graph Neural Network (GNN) inference is used in many real-world applications. 
Data sparsity in GNN inference, including sparsity in the input graph and the GNN model, offer opportunities to further speed up inference. Also, many pruning techniques have been proposed for model compression that increase the data sparsity of GNNs.

We propose Dynasparse, a comprehensive hardware-software codesign on FPGA to accelerate GNN inference through dynamic sparsity exploitation. For this, we decouple the GNN computation \emph{kernels} from the basic computation \emph{primitives}, and explore hardware-software codesign as follows:  1) \emph{Hardware design}: We propose a novel unified accelerator design on FPGA to efficiently execute various computation primitives. We develop a customized soft processor that is tightly coupled with the accelerator to execute a runtime system. Moreover, we develop  efficient hardware mechanisms to profile the data sparsity and perform on-the-fly data format transformation to prepare the input data for various computation primitives; 2)  \emph{Software design}: We develop a runtime system that works synergistically with the accelerator to perform dynamic kernel-to-primitive mapping based on data sparsity. We implement Dynasparse on a state-of-the-art FPGA platform, Xilinx Alveo U250, and evaluate the design using widely used GNN models (GCN, GraphSAGE, GIN and SGC). For the above GNN models and various input graphs,  the proposed accelerator and  dynamic kernel-to-primitive mapping reduces the inference latency by $3.73\times$ on the average compared with the static mapping strategies employed in the state-of-the-art GNN accelerators. Compared with state-of-the-art CPU (GPU) implementations, Dynasparse achieves up to $56.9\times$ ($2.37\times$)  speedup in end-to-end latency. Compared with state-of-the-art FPGA implementations, Dynasparse achieves $2.7\times$ speedup in accelerator execution latency.

\end{abstract}


\input{1-introduction}

\input{2-background}
\input{3-problem-definition}

\input{4-overview}

\input{compiler}

\input{5-hardware}

\input{6-runtime-system}
\input{7-analaysis}

\input{8-experiment}
\input{9-conclusion}

\bibliographystyle{IEEEtran}
\bibliography{reference}

\end{document}

%% file: 1-introduction.tex
\section{Introduction}

Graph Neural Networks (GNNs) have achieved great success in many real-world applications, such as recommendation systems, social media, etc.
\emph{Low-latency GNN inference} is needed in many real-world applications, such as  traffic prediction \cite{zhong2022explainable}, scientific simulation \cite{hewes2021graph}, etc.

While many techniques \cite{yan2020hygcn, zhang2021boostgcn, zhang2020hardware, lin2022hp,  meng2021dynamap, zhang2023graphagile, sarkar2022flowgnn} have been proposed to accelerate GNN inference, no work has systematically studied the data sparsity in GNNs to reduce the inference latency. GNNs (\cite{kipf2016semi, hamilton2017inductive}) involve various computation kernels, where there are three types of data sparsity: (1) \emph{Sparsity of graph structure}: The graphs in the real-world applications are usually sparse, as most vertices have a small number of neighbors, (2) \emph{Sparsity of vertex features}: The vertex features have various sparsity depending on the property of the graphs, activation function, etc., and (3) \emph{Sparsity of GNN model}: The weight matrices in GNN models can also have data sparsity due to model pruning, etc. Moreover, the data sparsity can vary significantly based on the input graphs and GNN models (See Section \ref{subsec:GNN-sparsity}).  Prior works directly map the GNN kernels to computation primitives (See Section \ref{subsec:GNN-sparsity}), and do not consider data sparsity, leading to potentially suboptimal performance. 

To efficiently utilize the data sparsity in GNN inference, we propose to decouple the GNN \emph{kernels} (feature aggregation and feature transformation) from the basic \emph{primitives} (dense-dense matrix multiplication (GEMM), sparse-dense matrix multiplication (SpDMM),  sparse-sparse matrix multiplication (SPMM)). A GNN kernel can be dynamically mapped to a primitive according to the sparsity of the data. 
However, there are several challenges: (1) While the sparsity of the graph structure and GNN model is known before the execution of inference (runtime), the sparsity of vertex features in the intermediate layers is  known only at runtime. Therefore, static (compile time) kernel-to-primitive mapping may not be optimal. (2) While the GNN kernels can be mapped to various primitives, these primitives have different data formats and layouts. Switching the data format and the data layout can incur large overhead during execution. (3) Different primitives have different computation patterns and memory access patterns. While general purpose processors are efficient for dense primitives (GEMM), their data path and cache organization are inefficient for sparse primitives (SpDMM, SPMM). 

To address the above challenges, we propose Dynasparse, a hardware-software codesign, which can efficiently exploit the data sparsity in GNN inference. For the hardware design, we use Field Programmable Gate Array (FPGA) as the target hardware platform. The programmability of FPGA allows us to (1) develop a customized data path and memory organization to support various computation primitives, (2) develop efficient hardware mechanism for sparsity profiling and transformation of data format and data layout (Section \ref{subsec:data-format}), and (3) implement a lightweight and customized soft processor to perform dynamic kernel-to-primitive mapping at runtime. We summarize our main contributions as follows:

\begin{itemize}
    { 
    \item We develop a complete system on FPGA with the following innovations in hardware design: 
    \begin{itemize}
        \item a novel hardware architecture, named Agile Computation Module, consisting of multiple Computation Cores with flexible data path and memory organization that can execute various computation primitives, including GEMM, SpDMM and SPMM.
        \item an efficient hardware mechanism that supports fast sparsity profiling and data format/layout transformation.
    \end{itemize} 
    \item We propose a soft processor and develop a runtime system on the soft processor to enable dynamic sparsity exploitation, including:
    \begin{itemize}
        \item dynamic kernel-to-primitive (K2P) mapping strategy that automatically selects the optimal computation primitive for a given kernel based on an analytical performance model.
        \item task scheduling strategy that manages the execution of the computation primitives on the accelerator to achieve load balance across multiple Computation Cores in the FPGA accelerator.
    \end{itemize}}
    \item We implement the proposed codesign on a state-of-the-art FPGA, Xilinx Alveo U250. For various GNN models and input graphs,  the proposed accelerator and the dynamic kernel-to-primitive mapping reduce the inference latency by $3.73\times$ on the average compared with the static mapping strategies employed in the state-of-the-art GNN accelerators. { Compared with state-of-the-art CPU (GPU) implementations, Dynasparse achieves up to $56.9\times$ ($2.37\times$)  speedup in end-to-end latency. Compared with state-of-the-art FPGA implementations, Dynasparse achieves $2.7\times$ speedup in accelerator execution latency.}
\end{itemize}

%% file: 2-background.tex
\section{Background}

\subsection{Graph Neural Network}

 GNNs \cite{kipf2016semi, hamilton2017inductive} are proposed for representation learning on graphs $  \mathcal{G}(\mathcal{V},\mathcal{E})$, and 
 follow the  message-passing paradigm (Algorithm \ref{alg:GNN-computation-abstraction}) in which the vertices recursively aggregate information from the neighbors. $\bm{h}^{L}_{v}$ denotes the last-layer embedding of the target vertex $v$. The Update() is usually a Multi-Layer Perceptron that transforms the vertex features.  
An element-wise activation function is applied to the feature vectors after the Aggregate() and Update() in each layer. 
 The output embedding $\bm{h}^{L}_{v}$
 can be used for many downstream tasks, such as node classification (\cite{hamilton2017inductive,kipf2016semi}), link prediction, etc. GCN \cite{kipf2016semi}, GraphSAGE \cite{hamilton2017inductive}, GIN \cite{xu2018powerful}, and SGC \cite{wu2019simplifying} are some representative GNN models. Table \ref{tab:notations} summarizes the notations used in this paper.
 


\begin{table}[h]
\centering
\caption{Notations}
\begin{adjustbox}{max width=0.48\textwidth}
\begin{tabular}{cc|cc}
\toprule
 \textbf{{Notation}} & \textbf{{Description}}  & \textbf{{Notation}}  & \textbf{{Description}} \\
 \midrule
\midrule
{$  \mathcal{G}(\mathcal{V},\mathcal{E})$ }& {input graph}  &  $ v_{i}$ & {$i^{\text{th}}$ vertex} \\ \midrule
$ \mathcal{V}$ &  {set of vertices} &  $ e_{ij}$ & {edge from $ v_{i}$ to $  v_{j}$} \\ \midrule
$ \mathcal{E}$& {set of edges} &  $ L$&{number of GNN layers} \\ \midrule
$\bm{A}$& graph adjacency matrix &  $ \mathcal{N}(i)$& the set of neighbors of $ v_{i}$ \\ \midrule
$ \bm{h}_{i}^{l-1}$& input feature vector of $ v_{i}$
at layer $l$    & $\bm{W}^{l}$  &  weight matrix of layer $l$  \\  \midrule
$\bm{H}^{l-1}$ & input feature matrix to layer $l$ & $\sigma()$ & activation function \\
 \bottomrule
\end{tabular}
\end{adjustbox}
\label{tab:notations}
\end{table}

\begin{algorithm}
\caption{GNN Computation Abstraction}
\label{alg:GNN-computation-abstraction}
\begin{small}
\begin{algorithmic}[1]
 \renewcommand{\algorithmicrequire}{\textbf{Input:}}
\renewcommand{\algorithmicensure}{\textbf{Output:}}
 \Require Input graph: $\mathcal{G}(\mathcal{V},\mathcal{E})$; vertex features: $\left\{\bm{h}^{0}_{1}, \bm{h}^{0}_{2}, \bm{h}^{0}_{3}, ..., \bm{h}^{0}_{|\mathcal{V}|}\right\}$;
 \Ensure Output vertex features $\left\{\bm{h}^{L}_{1}, \bm{h}^{L}_{2}, \bm{h}^{L}_{3}, ..., \bm{h}^{L}_{|\mathcal{V}|}\right\}$;
\For{$l=1...L$}
\For{each vertex $v \in \mathcal{V}$}
\State{$\bm{a}^l_{v} = {\text{Aggregate}(}\bm{h}_{u}^{l-1}: u\in \mathcal{N}(v))$}
\State{$\bm{z}_{v}^l = {\text{Update}(}\bm{a}_{v}^{l}, \bm{W}^{l} \textbf{)}$, $ \bm{h}_{v}^l = \sigma(\bm{z}_{v}^l )$}
\EndFor
\EndFor
\end{algorithmic}
\end{small}
\end{algorithm}


\subsection{Data Sparsity in GNN inference}
\label{subsec:GNN-sparsity}

\begin{figure}[ht]
     \centering
     \includegraphics[width=8.5cm]{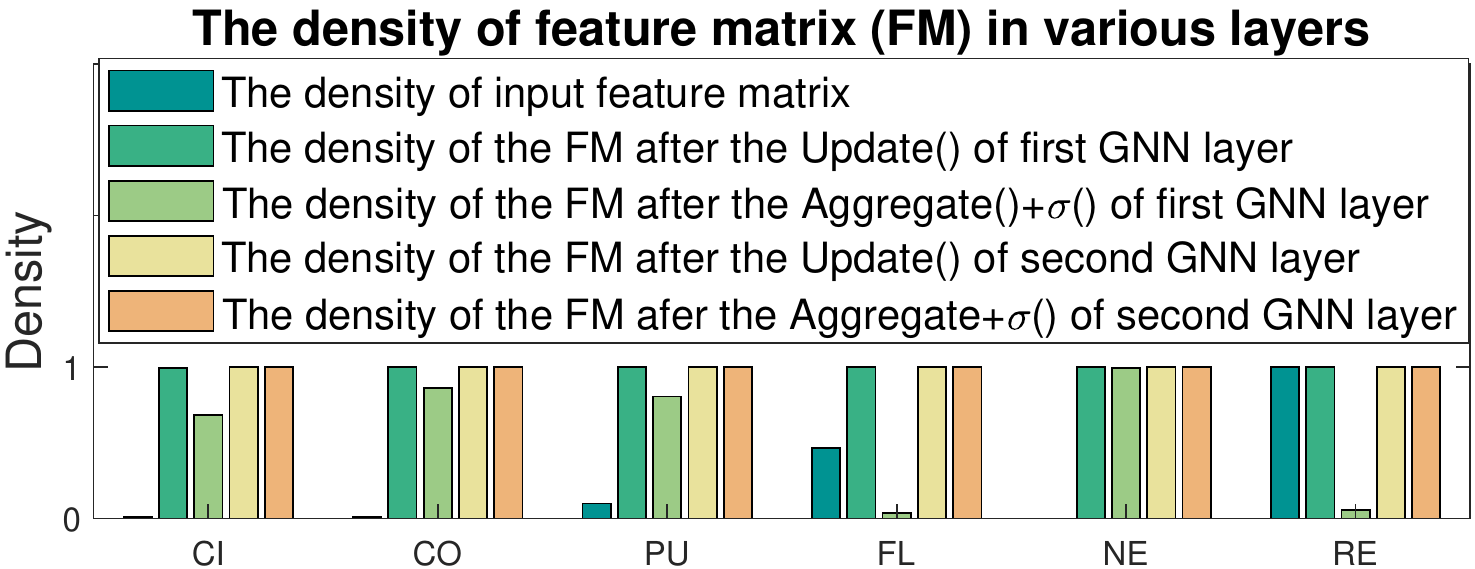}
     \caption{Density of the feature matrices in the GCN model \cite{kipf2016semi}}
     \label{fig:density-of-feature-matrix}
     \vspace{-0.2cm}
\end{figure}

The \emph{density} of a matrix is defined as the total number of non-zero elements divided by the total number of elements. Note that, the \emph{sparsity} is given by $(1 - \text{\emph{density}})$. The computation kernels in GNNs involve three types of matrices: graph adjacency matrix $\bm{A}$, vertex feature matrix $\bm{H}$, and weight matrix $\bm{W}$. 
The adjacency matrix $\bm{A}$ of different graph datasets \cite{pyg-dataset} can have different densities. For a given adjacency matrix, different parts of the matrix have different densities. Figure \ref{fig:density-of-feature-matrix}
shows the densities of feature matrices in GCN \cite{kipf2016semi}. For different graphs, the input feature matrices have different densities. The feature matrices of different layers also have different densities.  For the weight matrices, prior works (\cite{rahman2022triple, chen2021unified}) have  proposed various pruning techniques to reduce the density of the weight matrices. 

\begin{figure}[h]
     \centering
     \includegraphics[width=4cm]{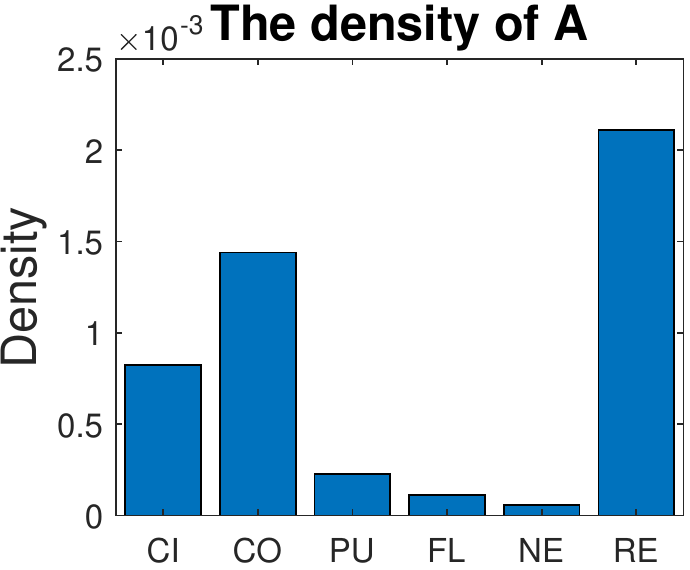}
     \includegraphics[width=4cm]{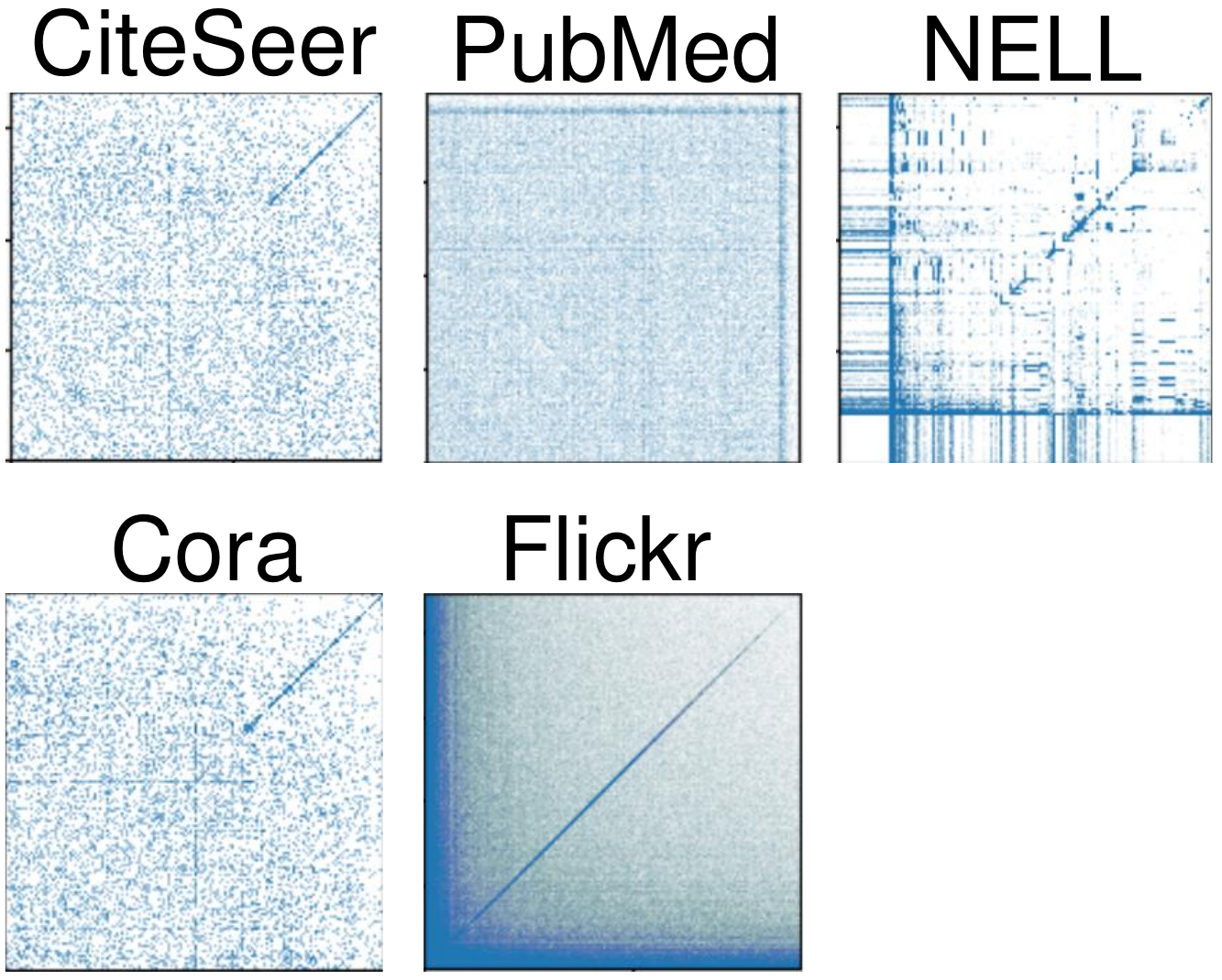}
     \caption{The density and the visualization of graph adjacency matrix $\bm{A}$ of various graphs \cite{pyg-dataset}}
     \label{fig:density-of-adjacency-matrix}
\end{figure}

\begin{figure*}[h]
     \centering
     \includegraphics[width=18cm]{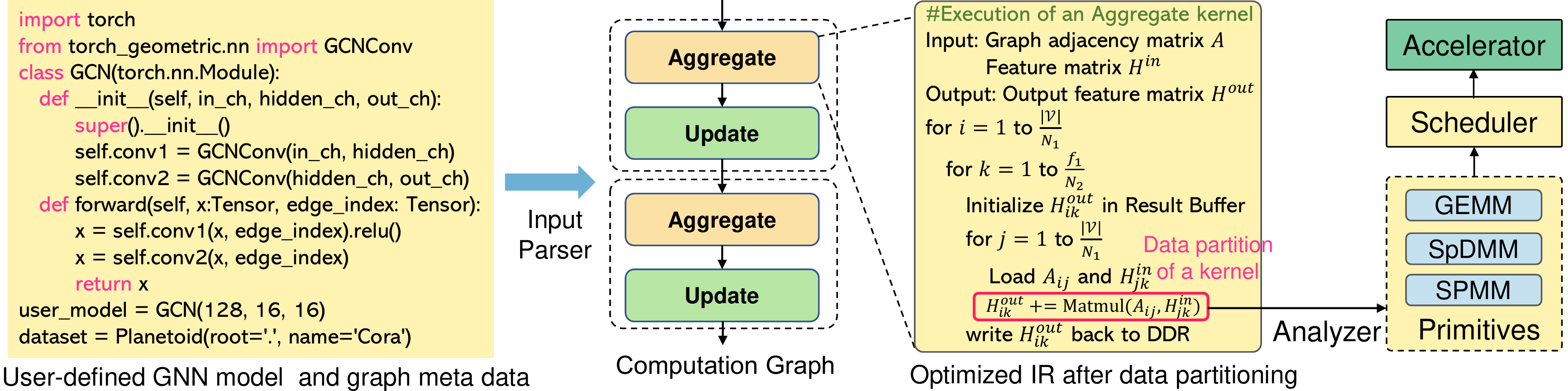}
     \caption{Proposed workflow}
     \label{fig:workflow}
\end{figure*}

\subsection{GNN Acceleration based on Data Sparsity}

Although there are various data sparsities in GNNs, no prior work has systematically studied exploiting the data sparsity for GNN inference acceleration.
HyGCN \cite{yan2020hygcn} and BoostGCN \cite{zhang2021boostgcn}  map Aggregate() to SpDMM and map update() to GEMM, ignoring the data sparsity in feature matrices and weight matrices.  AWB-GCN \cite{geng2020awb} maps both Aggregate() and update() to SpDMM. Then, they propose an accelerator to efficiently execute SpDMM. However, they do not exploit the data sparsity in weight matrices.  DeepBurning-GL \cite{liang2020deepburning} is a design automation framework that generates the optimized hardware accelerator given the information of the input graph and the GNN model. However, their framework needs to regenerate the optimized accelerator if the sparsity of the data is changed.  To summarize, prior GNN accelerators do not fully exploit the data sparsity in GNNs, or are not flexible to exploit data sparsity in GNN inference.


%% file: 3-problem-definition.tex
\section{Overview}

\subsection{Problem Definition}
\label{subsec:problem-define}

The computation \textbf{kernels} in GNN inference are feature aggregation and feature transformation which correspond to  Aggregate() and Update() in the message-passing paradigm of GNN (Algorithm \ref{alg:GNN-computation-abstraction}). 
\begin{itemize}
    \item  Aggregate(): The input is graph adjacency matrix $\bm{A}$ and  feature matrix $\bm{H}_{\text{in}}$. The output is $\bm{H}_{\text{out}} = \bm{A} \times\bm{H}_{\text{in}}$.
    \item  Update(): The input is vertex feature matrix  $\bm{H}_{\text{in}}$ and weight matrix $\bm{W}$. The output is  $\bm{H}_{\text{out}} = \bm{H}_{\text{in}} \times \bm{W}$.
\end{itemize}
The computation \textbf{primitives} are GEMM, SpDMM and SPMM. While all  the primitives perform multiplication of two input matrices to produce an output matrix, they have different ways of dealing with the zero elements: (1) GEMM views the two input matrices as dense matrices, and performs multiply-accumulate for all the matrix elements no matter whether an element is non-zero or not. (2) SpDMM views one input matrix as sparse matrix and skips the computation operations for all the zero elements in this input matrix. (3) SPMM takes two input sparse matrices and skips the computation operations for all the zero elements in the two input matrices. 



This work targets full-graph inference: given a GNN model and an input graph, we perform the message-passing paradigm (Algorithm \ref{alg:GNN-computation-abstraction}) in the full input graph to obtain the embeddings of all the vertices. Full-graph inference has been widely studied in the literature \cite{yan2020hygcn, geng2020awb, zhang2021boostgcn}. Our objective is to exploit the data sparsity of GNN kernels to further accelerate the inference process. We assume that the sparsity of the data is unknown before the accelerator design or hardware execution. Our intent is to develop a single hardware-software codesign on FPGA that is efficient and flexible to support  various graphs and GNN models of various data sparsity. Therefore, the proposed work does not require regenerating the FPGA accelerator if data sparsity changes.



%% file: 4-overview.tex
\subsection{System Overview}

\begin{figure}[h]
     \centering
     \includegraphics[width=8.5cm]{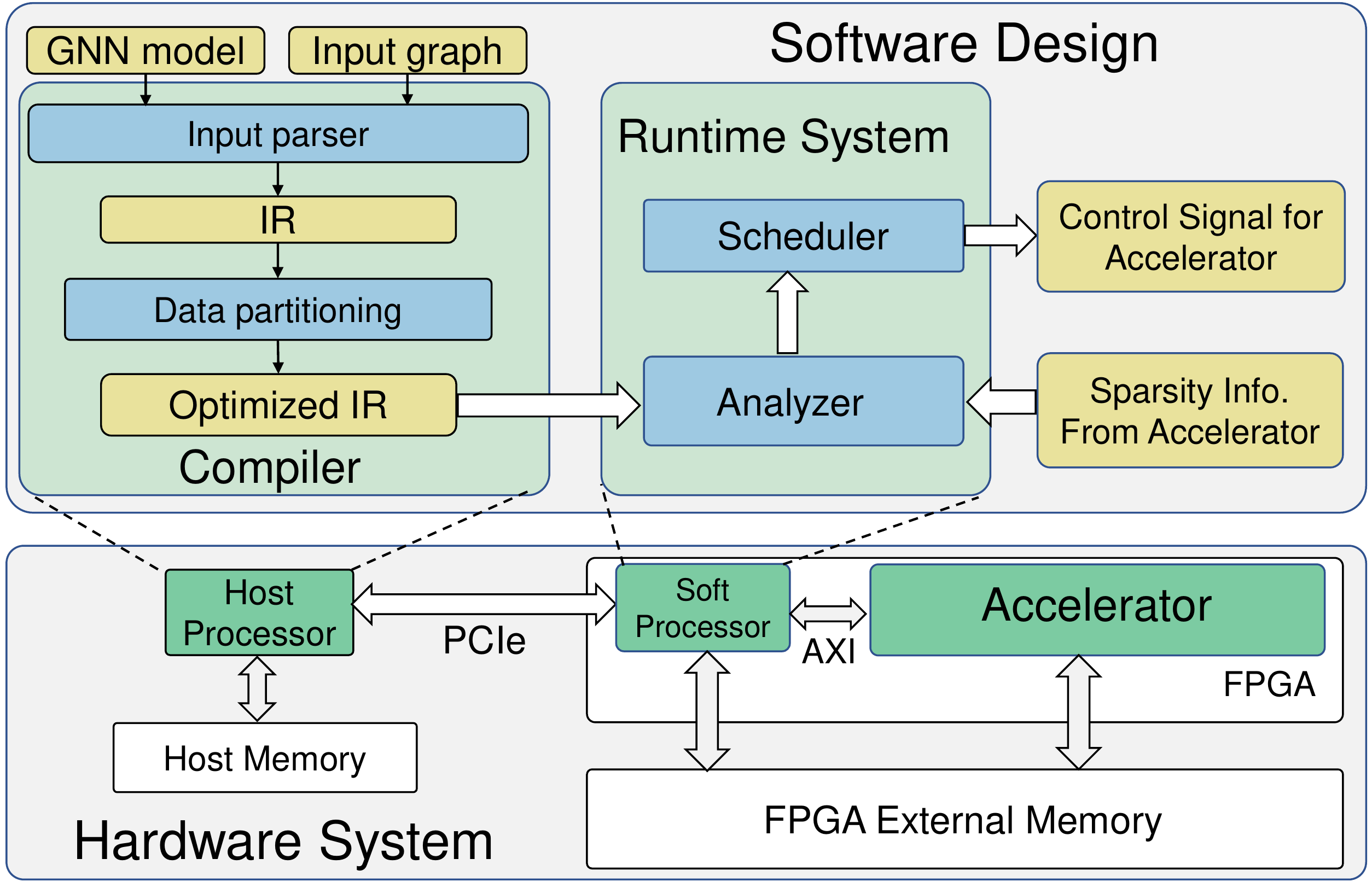}
     \caption{Overview of the proposed system}
     \label{fig:system-overview}
\end{figure}

 {
Figure \ref{fig:system-overview} depicts the proposed system design. The software comprises of a \emph{compiler} and a \emph{runtime system}. The hardware system has three components:
\begin{itemize}
    \item \textbf{Host processor}: The compiler is executed on the host processor to perform compilation (preprocessing) for the input GNN model and the input graph to generate the intermediate representation (IR). The IR is sent to the soft processor for execution.
    \item \textbf{Soft Processor on FPGA}: The runtime system is executed on the soft processor. It takes the IR as input, and dynamically schedules the computation tasks on the accelerator by sending control signals to the accelerator. 
    \item \textbf{Accelerator on FPGA}: It executes the three computation primitives (GEMM, SpDMM, SPMM), profiles data sparsity, and performs data layout/format transformation. It receives the control signals from the soft processor to execute the computation tasks, and also sends the data sparsity information to the soft processor at runtime.
\end{itemize}
}

{
The workflow is illustrated in Figure \ref{fig:workflow}. The execution of GNN inference consists of two steps: }

 {
\vspace{0.1cm}
\noindent \textbf{Step 1. Compilation/Preprocessing}: The compiler (See Section \ref{sucsec:IR}) performs the following preprocessing: \circled{1} \textbf{Generating intermediate representation (IR):} It takes the specifications of the user-defined GNN model and the graph meta data as input, and generates the IR for the GNN computation graph (See Figure \ref{fig:workflow}).
\circled{2} \textbf{Data partitioning}: The compiler performs data partitioning for each kernel. Data partitioning is required since (1) in real-world applications, the input graph can be very large and the FPGA accelerator has limited on-chip memory, (2) within a matrix,
different parts of the matrix can have different data sparsity. Data partitioning enables fine-grained kernel-to-primitive mapping (See Section \ref{subsec:dyna-k2p-mapping}), leading to more efficient sparsity exploitation. \circled{3} \textbf{Preprocessing of data sparsity}: When the compiler performs data partitioning, it uses counters to profile the sparsity information of graph adjacency matrix $\bm{A}$, weight matrix $\bm{W}$, and input feature matrix  $\bm{H}^{0}$. Note that the sparsity information of the feature matrices in the intermediate layers $\{\bm{H}^{1},...,\bm{H}^{L}\}$ is unknown at compile time and is profiled by the accelerator at runtime.}




{
\vspace{0.1cm}
\noindent \textbf{Step 2. Runtime execution}: At runtime, the soft processor and the accelerator collaborate to perform GNN inference. The runtime system on the soft processor consists of \emph{an Analyzer} and \emph{a Scheduler}. The accelerator contains multiple Computation Cores. The Analyzer takes the optimized IR from the compiler and the data sparsity information from the compiler and the accelerator to dynamically map a kernel to a primitive based on a performance model. Then, the Scheduler schedules the execution of the primitives on the accelerator (Section \ref{subsec:task-scheduling}). The runtime system performs dynamic kernel-to-primitive (K2P) mapping. Note that the mapping must be performed dynamically at runtime: (1) The densities of the feature matrices in the intermediate layers $\{\bm{H}^{1},...,\bm{H}^{L}\}$  are unknown before runtime; (2) The Computation Core has various execution modes (Section \ref{subsec:Microarchitecture}) with each mode executing a specific primitive. These execution modes have different computation efficiency (See Section \ref{subsec:performance-model}) with respect to the density of data. As a result, for a computation kernel of high density, executing it using GEMM primitive on the Computation Core will be more efficient. For a GNN kernel of low density, executing it using SpDMM or SPMM primitive on  the Computation Core will be more efficient. To handle this scenario, we build an analytical performance model (Section \ref{subsec:dyna-k2p-mapping}) to estimate the execution latency of a given primitive on the Computation Core with respect to the data sparsity. }


\vspace{0.1cm}
 {
The rest of the paper is organized as follows:  Section \ref{sec:compiler} covers the details of the compiler; Section \ref{sec:accelerator-design} introduces the proposed accelerator design; Section \ref{sec:runtime-system} introduces the proposed runtime system; Section \ref{sec:Implementation details} and \ref{sec:Evaluation-Results} describe the implementation details and evaluation results, respectively.
 }


%% file: compiler.tex
\section{Compiler}
\label{sec:compiler}

\subsection{Intermediate Representation (IR)}
\label{sucsec:IR}
 {
We define the meta data in the IR in Table \ref{tab:inter-representation}, including the meta data of the kernel and the meta data of the execution scheme. The execution scheme of a kernel is the plan for executing the kernel. The IR defines two types of kernels -- \emph{Aggregate} and \emph{Update}, corresponding to  $\text{Aggregate}()$ and $\text{Update}()$ in the GNN abstraction (See Algorithm \ref{alg:GNN-computation-abstraction}). 
}



\begin{table}[ht]
\centering
\caption{Meta data of a kernel in the IR }
\begin{adjustbox}{max width=0.55\textwidth}
\begin{tabular}{|l|l|}
\hline
\textbf{Layer Type} & \begin{tabular}[|c|]{@{}c@{}} Aggregate(0), Update(1) \end{tabular} \\ 
\hline
\textbf{Layer ID}  & 1,2,3,... \\
\hline
\textbf{Input Dimension}  & $f_{\text{in}}$  \\
\hline
\textbf{Output Dimension}  & $f_{\text{out}}$ \\
\hline
\textbf{\# of vertices} & $|\mathcal{V}|$  \\
\hline
\textbf{\# of edges} & $|\mathcal{E}|$  \\
\hline
\textbf{Aggregation operator} &  Max, Sum, Min, Mean \\
\hline
\textbf{Activation type} & ReLU, PReLU\\
\hline
\textbf{Activation enabled} & True, False\\
\hline
\textbf{Meta data of execution scheme} & \{...\} (See Algorithm \ref{alg:scheme-aggregate} and \ref{alg:scheme-Update})\\
\hline
\end{tabular}
\end{adjustbox}
\label{tab:inter-representation}
\end{table}






\begin{figure}[h]
     \centering
     \includegraphics[width=8.5cm]{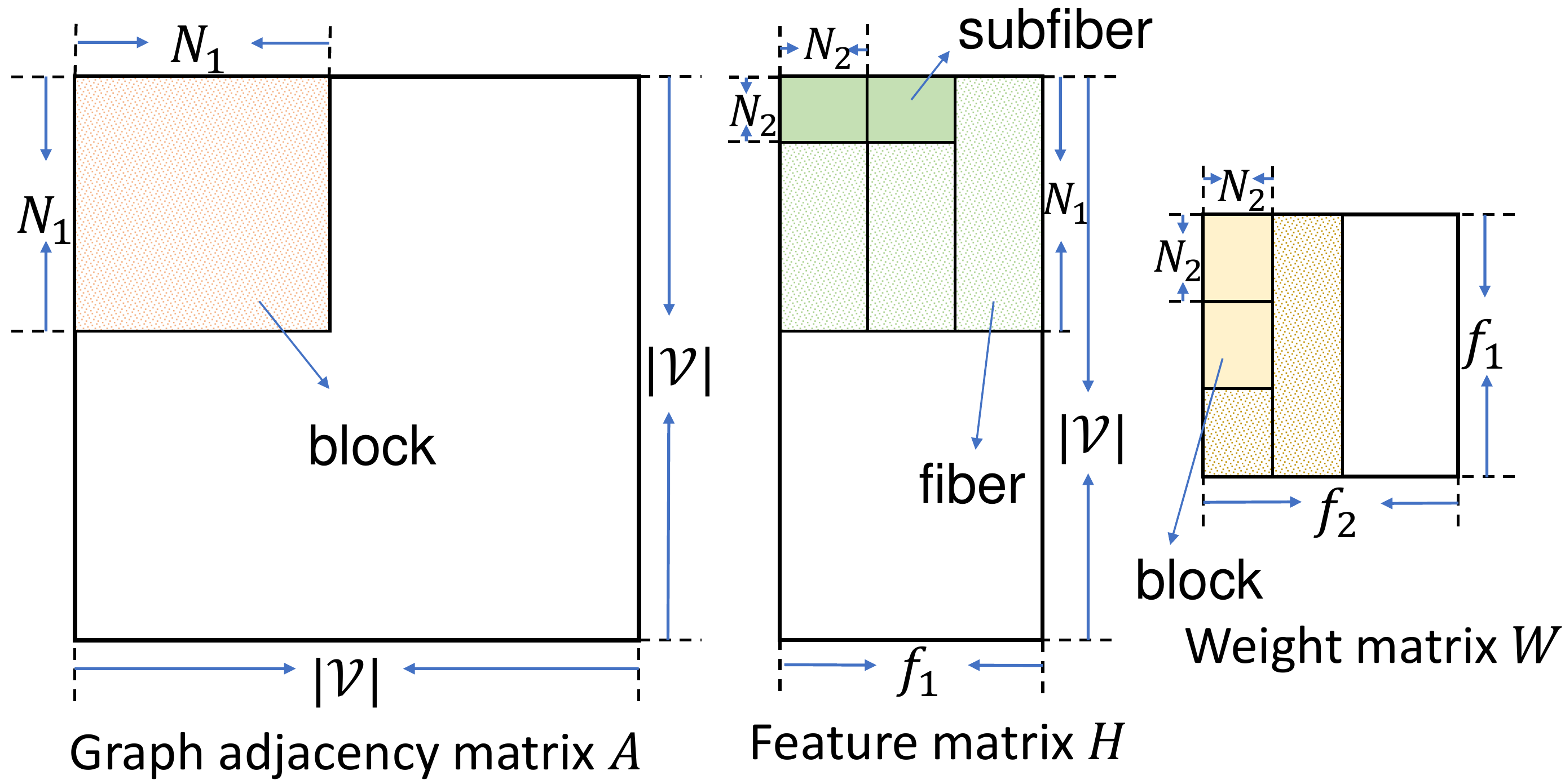}
     \caption{Illustration of data and model partitioning}
     \label{fig:Partition-scheme}
\end{figure}

 {
\subsection{Compilation Process}
The compilation process has two steps (See Figure \ref{fig:system-overview}):
\begin{itemize}
    \item \textbf{Step 1 (parsing the input)}: The compiler takes the specification of the GNN model (Defined using Pytorch Geometric Library \cite{pyg-dataset}) and the graph meta data as input, and generates the computation graph for GNN inference (See the example in Figure \ref{fig:workflow}).  The computation graph has $\sum_{l=1}^{L} k_{l}$ nodes, where $L$ denotes the number of GNN layers in the GNN model and $k_{l}$ denotes the number of kernels in layer $l$ ($1 \leqslant l \leqslant L$). In the computation graph, each node represents the IR of a kernel. An edge denotes the data dependency between two kernels.
    \item \textbf{Step 2 (data partitioning and execution scheme generation)}: The compiler performs data partitioning for each kernel and generates the execution scheme for the kernel. Then, the meta data of the execution scheme is stored in the IR to produce the  optimized IR (See Figure \ref{fig:workflow}) that is sent to the runtime system. 
\end{itemize} 
}

\subsection{Data Partitioning}
\label{subsec:data-partitoning}

Figure \ref{fig:Partition-scheme} depicts the proposed data partition scheme. The graph adjacency matrix $\bm{A}$ has the dimension $|\mathcal{V}| \times |\mathcal{V}|$. $\bm{A}$ is partitioned into blocks with each block having dimension of $N_{1} \times N_{1}$. We use $\bm{A}_{ij}$ to denote a block where $\bm{A}_{ij} = \bm{A}[i*N_{1}:(i+1)*N_{1}][j*N_{1}:(j+1)*N_{1}]$. The feature matrix $\bm{H}$ of dimension $|\mathcal{V}| \times f_{1}$ is partitioned into fibers. Each fiber has  dimension $N_{1} \times N_{2}$ and  $\bm{H}_{ij} = \bm{H}[i*N_{1}:(i+1)*N_{1}][j*N_{2}:(j+1)*N_{2}]$.  We further partition each fiber into subfibers where each subfiber has size $N_{2} \times N_{2}$. $\bm{H}_{ij-k}$ denotes the $k^{\text{th}}$ subfiber of $\bm{H}_{ij}$. We use $\bm{H}_{i-k}$ to denote the concatenation of $\{\bm{H}_{i1-k}$, $\bm{H}_{i2-k}$, ..., $\bm{H}_{i\frac{N_{1}}{N_{2}}-k}\}$. The weight matrix $\bm{W}$ is partitioned into blocks with each block having size of $N_{2}\times N_{2}$. $\bm{W}_{ij} = \bm{W}[i*N_{2}:(i+1)*N_{2}][j*N_{2}:(j+1)*N_{2}]$.  



\subsection{Execution Scheme}
\label{subsec:execution-scheme}

Based on the  data partition scheme, the compiler generates the execution plan for each computation kernel, shown in Algorithm \ref{alg:scheme-aggregate} and \ref{alg:scheme-Update}. The execution of a computation \emph{kernel} is decomposed into a set of independent computation \emph{tasks}. Each task performs the execution of an output data partition and there is no data dependency among the tasks within a kernel. Each task performs the multiplication of data partitions to obtain an output data partition, and the computation  primitive to execute the matrix multiplication $\verb|Matmul()|$ is determined by the Runtime System. We generalize the representation of a task in Algorithm \ref{alg:Computation-Task}.

\begin{algorithm}
\caption{Execution scheme of an Aggregate kernel}
\label{alg:scheme-aggregate}
\begin{small}
\begin{algorithmic}[1]
 \renewcommand{\algorithmicrequire}{\textbf{Input:}}
\renewcommand{\algorithmicensure}{\textbf{Output:}}
 \Require Graph adjacency matrix $\bm{A}$; Input feature matrix $\bm{H}^{\text{in}}$;
 \Ensure Output  feature matrix $\bm{H}^{\text{out}}$;
\State{\tikzmark{start2} \hspace{-0.25cm} Execute the Aggregate kernel}
\For{$i=1$ to $\frac{|\mathcal{V}|}{N_{1}}$}
    \For{$k=1$ to $\frac{f_{1}}{N_{2}}$}
        \State{ \tikzmark{start1} \hspace{-0.35cm}  Initialize $\bm{H}^{\text{out}}_{ik}$ in the Result Buffer}
        \For{$j=1$ to $\frac{|\mathcal{V}|}{N_{1}}$}
            \State{ Load $\bm{A}_{ij}$ and $\bm{H}^{\text{in}}_{jk}$}
            \State{$\bm{H}^{\text{out}}_{ik} += \text{Matmul}(\bm{A}_{ij}, \bm{H}^{\text{in}}_{jk})$}
        \EndFor
        \State{Write $\bm{H}^{\text{out}}_{ik}$ back to DDR memory}
        \tikzmark{end1} \tikzmark{end2}
    \EndFor
\EndFor
\end{algorithmic}
\end{small}
\end{algorithm}
\vspace{-0.8cm}
\Textbox[2.0cm]{start1}{end1}{Task}
\Textbox[1.9cm]{start2}{end2}{Kernel}

\begin{algorithm}
\caption{Execution scheme of an Update Kernel}
\label{alg:scheme-Update}
\begin{small}
\begin{algorithmic}[1]
 \renewcommand{\algorithmicrequire}{\textbf{Input:}}
\renewcommand{\algorithmicensure}{\textbf{Output:}}
 \Require  Input  feature matrix $\bm{H}^{\text{in}}$; Weight matrix $\bm{W}$;
 \Ensure Output feature matrix $\bm{H}^{\text{out}}$;
 \State{\tikzmark{start4} \hspace{-0.25cm} Execute the Update kernel}
\For{$i=1$ to $\frac{|\mathcal{V}|}{N_{2}}$}
    \For{$k=1$ to $\frac{f_{2}}{N_{2}}$}
        \State{\tikzmark{start3} \hspace{-0.24cm}  $g = \lfloor\frac{i\times N_{2}}{N_{1}}\rfloor$, $f = i\%(\frac{N_{1}}{N_{2}})$}
        \State{Initialize $\bm{H}^{\text{out}}_{gk-f}$ in the Result Buffer}
        \For{$j=1$ to $\frac{f_{1}}{N_{1}}$}
            \State{ Load $\bm{H}^{\text{in}}_{gj-f}$ and $\bm{W}_{jk}$}
            \State{$\bm{H}^{\text{out}}_{gk-f} += \text{Matmul}(\bm{H}^{\text{in}}_{gj-f}, \bm{W}_{jk})$}
        \EndFor
        \State{Write $\bm{H}^{\text{out}}_{gk-f}$ back to DDR memory}
        \tikzmark{end3} \tikzmark{end4}
    \EndFor
\EndFor
\end{algorithmic}
\end{small}
\end{algorithm}
\vspace{-0.8cm}
\Textbox[2.0cm]{start3}{end3}{Task}
\Textbox[1.9cm]{start4}{end4}{Kernel}

\begin{algorithm}
\caption{A computation task}
\label{alg:Computation-Task}
\begin{algorithmic}[1]
 \renewcommand{\algorithmicrequire}{\textbf{Input:}}
\renewcommand{\algorithmicensure}{\textbf{Output:}}
 \Require  $\{\bm{X}_{i1}$, $\bm{X}_{i2}$, ..., $\bm{X}_{iK}\}$ and $\{\bm{Y}_{1j}$, $\bm{Y}_{2j}$, ..., $\bm{X}_{Kj}\}$;
 \Ensure Output matrix: $\bm{Z}_{ij}$;
\State{Initialize $\bm{Z}_{ij}$ in the Result Buffer}
\For{$k=1$ to $K$}
    \State{Load $\bm{X}_{it}$ and $\bm{Y}_{tj}$ onto the on-chip buffer}
    \State{$\bm{Z}_{ij} += \text{Matmul}(\bm{X}_{it}, \bm{Y}_{tj})$ }
\EndFor
\State{Write $\bm{Z}_{ij}$ back to DDR memory}
\end{algorithmic}
\end{algorithm}

%% file: 5-hardware.tex
\begin{figure*}[h]
     \centering
     \includegraphics[width=18cm]{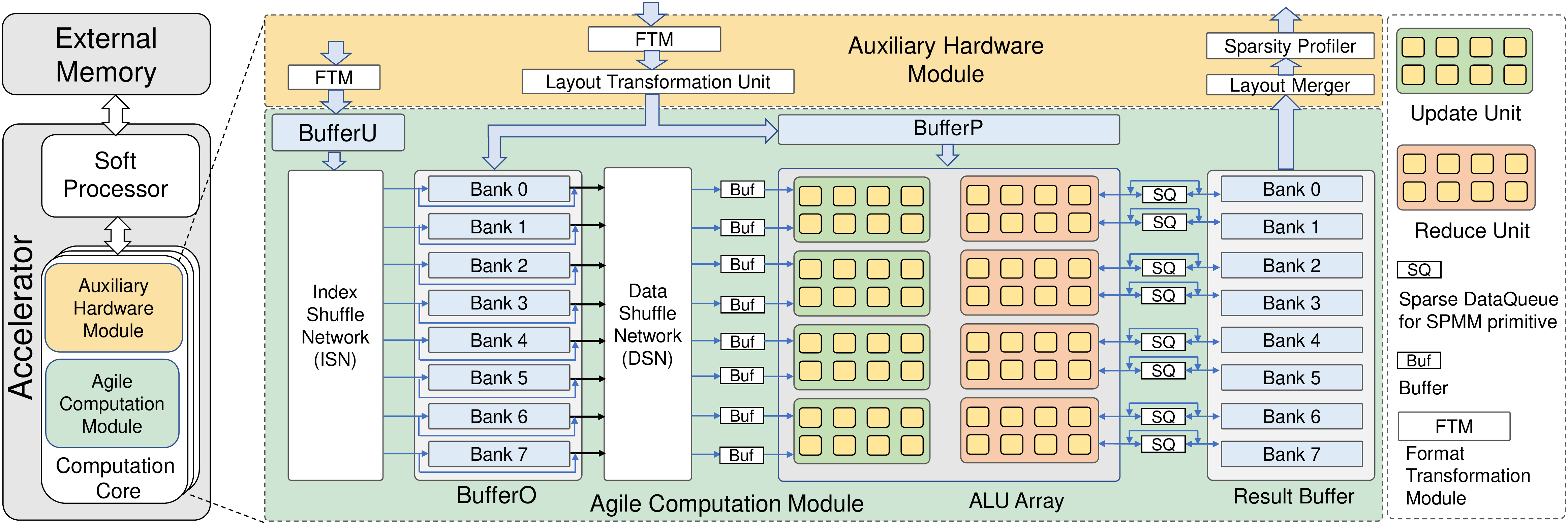}
     \caption{Diagram of a Computation Core}
     \label{fig:computation-core}
\end{figure*}

\begin{figure*}[h]
     \centering
     \includegraphics[width=18cm]{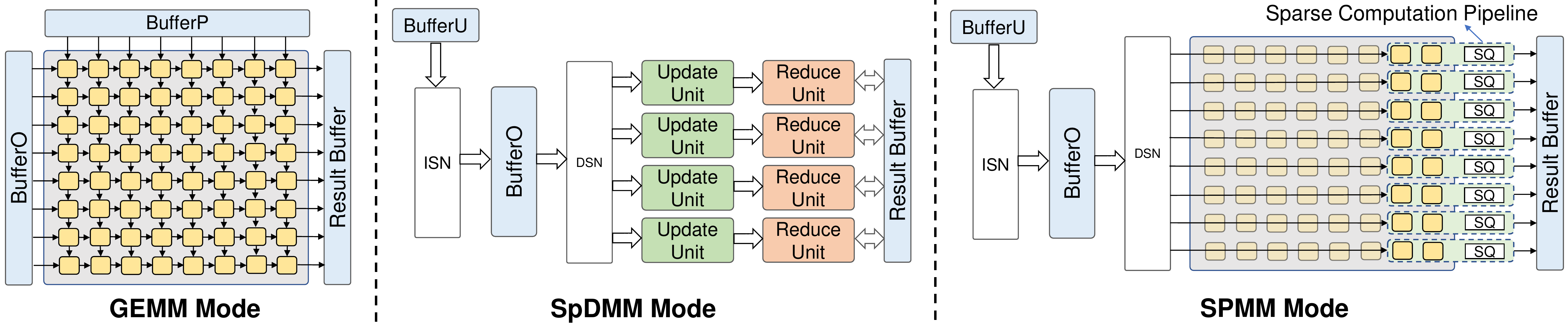}
     \caption{Various execution modes of  a Computation Core}
     \label{fig:execution-modes}
\end{figure*}

\section{Accelerator Design}
\label{sec:accelerator-design}
 {
 In Section \ref{subsec:data-format}, we introduce the data layout and data format that are used by Dynasparse. In Section \ref{subsubsec-ACM}, we introduce the Agile Computation Module which can execute three primitives (GEMM, SpDMM, and SPMM).  In Section \ref{subsubsec:AHM}, we describe the hardware mechanism for sparsity profiling, and data format/layout transformation.}

\subsection{Data format and data layout}
\label{subsec:data-format}
\noindent \textbf{Data format}: We store the matrices using \emph{sparse} format or \emph{dense} format. We use Coordinate (COO) format to represent a sparse matrix where  an nonzero element is represented using a three-tuple $(col, row, value)$ denoting the column index, row index, and value, respectively. COO format is the standard data format used in the state-of-the-art GNN libraries \cite{pyg-dataset}.

\vspace{0.1cm}
\noindent \textbf{Data layout}: It defines the order of storing the matrix elements. For a sparse matrix in the row-major order, the elements within the same $row$ are stored in contiguous locations. Otherwise, it is \emph{column-major} order. Similarly, row-major and column-major order for a dense matrix can be derived.

\vspace{0.1cm}
\noindent \textbf{Notations}: For a matrix $\bm{B}$, we use $\bm{B}[i]$ to denote the $i^{\text{th}}$ row of $\bm{B}$  and use  $\bm{B}[i:j]$ to denote the submatrix of $\bm{B}$ from $i^{\text{th}}$ row to $(j-1)^{\text{th}}$ row.  We use $\bm{B}[i][j]$ to denote the element of $\bm{B}$ at the $i^{\text{th}}$ row  and the $j^{\text{th}}$ column. An element $(j, i, value)$ in sparse $\bm{B}$ will also be denoted as $\bm{B}[i][j]=value$.


\subsection{Microarchitecture}
\label{subsec:Microarchitecture}
Each Computation Core (Figure \ref{fig:computation-core}) has an Agile Computation Module (ACM) and an Auxiliary Hardware Module (AHM). The ACM has an ALU (Arithmetic Logic Unit) array of dimension $p_{sys}\times p_{sys}$ and the interconnection among the ALUs are shown in Figure \ref{fig:execution-modes}. AHM performs sparsity profiling, data layout and format transformation (Section \ref{subsubsec:AHM}).

\subsubsection{Agile Computation Module (ACM)} \label{subsubsec-ACM} It has four data buffers -- BufferU, BufferO, BufferP and Result Buffer (RB). Buffer[U/O/P] store the input matrices and RB stores the output matrix. Each Buffer has $p_{sys}$ memory banks (denoted bank $0$ to bank $p_{sys} - 1$) for parallel on-chip memory access.  Each ALU can execute various arithmetic operations, including multiplication, max, addition, etc. 
There are two interconnection networks -- Index Shuffle Network (ISN) and Data Shuffle Network (DSN) -- for data communication.
The ACM has three execution modes -- \emph{GEMM mode}, \emph{SpDMM mode} and \emph{SPMM mode}.  The required data format and layout for various execution modes are summarized in Table \ref{tab:my_label}.

\begin{table}[]
\centering
\caption{Buffer (data format) [data layout] requirement to store the input/output matrices for executing $\bm{Z} = \bm{X} \times \bm{Y}$ in the three execution modes} 
\begin{adjustbox}{max width=0.48\textwidth}
\begin{tabular}{cccc}
 \toprule
  &  $\bm{X}$ &  $\bm{Y}$ & $\bm{Z}$\\ 
 \midrule
 \midrule
 GEMM &  \begin{tabular}[c]{@{}c@{}}  BufferO (dense) \\ \text{[}row major\text{]} \end{tabular} 
 & \begin{tabular}[c]{@{}c@{}} BufferP (dense) \\ \text{[}column major\text{]} \end{tabular}  & 
 \begin{tabular}[c]{@{}c@{}} Result Buffer (dense) \\ \text{[}row major\text{]} \end{tabular} \\  \midrule
 SpDMM & \begin{tabular}[c]{@{}c@{}} BufferU (sparse) \\ \text{[}row or column major\text{]} \end{tabular} 
 & \begin{tabular}[c]{@{}c@{}} BufferO (dense) \\ \text{[}row major\text{]} \end{tabular}  & 
 \begin{tabular}[c]{@{}c@{}} Result Buffer (dense) \\ \text{[}row major\text{]} \end{tabular}  \\  \midrule
 SPMM
 & \begin{tabular}[c]{@{}c@{}} BufferU (sparse) \\ \text{[}row major\text{]} \end{tabular} 
 & \begin{tabular}[c]{@{}c@{}} BufferO (sparse)   \\ \text{[}row major\text{]} \end{tabular} 
 & \begin{tabular}[c]{@{}c@{}}  Result Buffer (dense) \\ \text{[}row major\text{]} \end{tabular}  \\ \bottomrule
 \hline
\end{tabular}
\end{adjustbox}
\label{tab:my_label}
\end{table}

\vspace{0.1cm}
\noindent \textbf{GEMM Mode}: The ALU array is organized as a two-dimensional systolic array (See Figure \ref{fig:execution-modes}) to execute GEMM using output stationary dataflow. 
The systolic array can execute $p_{sys}^2$ multiply-accumulate (MAC) operations per clock cycle.


\begin{algorithm}
\caption{SpDMM using Scatter-Gather Paradigm}\label{alg:scatter-gather}
\begin{small}
\begin{algorithmic}[1]
\renewcommand{\algorithmicrequire}{\textbf{Input:}}
\renewcommand{\algorithmicensure}{\textbf{Output:}}
 \Require Sparse matrix (BufferU): $\bm{X}$; Dense matrix (BufferO): $\bm{Y}$;
 \Ensure Output matrix  (Result Buffer): $\bm{Z}$ ($\bm{Z} = \bm{X}\times \bm{Y}$);
\While {not done}
\For{each $e(i,j,value)$ in $\bm{X}$ \textbf{Parallel}}  {\color{blue}\Comment{Scatter Phase}}
\State Fetch $\bm{Y}[i]$ from BufferO   {\color{blue}\Comment{ISN routes $e$ to BufferO}}
\State Form input pair ($\bm{Y}[i]$, $e$)  
\State {\color{blue} \# DSN routes input pair to Update Units}
\EndFor

\For{each input pair \textbf{Parallel}}
{\color{blue}\Comment{Gather Phase}}
\State $u \gets ${Update($\bm{Y}[i],e.value$)}  {\color{blue}\Comment{Update Unit}}
\State Fetch $\bm{Z}[j]$ from Result Buffer
\State {$\bm{Z}[j] \gets$ {Reduce($u$)}}   {\color{blue}\Comment{Reduce Unit}}
\EndFor
\EndWhile
\end{algorithmic}
\end{small}
\end{algorithm}

\noindent \textbf{SpDMM Mode}: The ALU array is divided into $p_{sys}/2$ Update Units and $p_{sys}/2$ Reduce Units. Each Update or Reduce Unit has an ALU array of size $p_{sys}/2 \times 2$. Multiplication of a sparse matrix with a dense matrix is executed using the Scatter-Gather Paradigm shown in Algorithm \ref{alg:scatter-gather}. The sparse matrix denoted as $\bm{X}$ (in BufferU) is stored in row-major order using COO format. The dense matrix   denoted as $\bm{Y}$ (in BufferO) is stored in row-major order using dense format, and $\bm{Y}[i]$ is stored in bank $(i\mod p_{\text{sys}})$ of BufferO. Each non-zero element $e(i,j,weight)$  in  $\bm{X}$ is fetched from the BufferU ($p_{sys}/2$ elements can be fetched from BufferU per cycle) and sent to the ISN. Then $e$ is routed to bank $(i\mod p_{\text{sys}})$  for fetching  $\bm{Y}[i]$, which forms the input data pair ($\bm{Y}[i]$, $e$). The input pair is routed to the $(j\mod p_{\text{sys}}/2)^{\text{th}}$ Update Unit. The Update Unit performs the multiplication of $e.value$ and $\bm{Y}[i]$ to produce the intermediate result $u$. Then the corresponding Reduce Unit adds $u$ to $\bm{Z}[j]$.  SpDMM Mode can efficiently skip zero elements in the sparse matrix $\bm{X}$. The SpDMM Mode can execute $p_{sys}^{2}/2$ MAC operations per clock cycle.


\begin{algorithm}
\caption{SPMM using Row-wise Product with Scatter-Gather Paradigm}\label{alg:scatter-gather-SPMM}
\begin{small}
\begin{algorithmic}[1]
\renewcommand{\algorithmicrequire}{\textbf{Input:}}
\renewcommand{\algorithmicensure}{\textbf{Output:}}
 \Require  Sparse matrix (BufferU): $\bm{X}$; Sparse matrix (BufferO): $\bm{Y}$;
 \Ensure Output matrix (In Result Buffer): $\bm{Z} = \bm{X}\times \bm{Y}$;
\For{each row $\bm{Z}[j]$ in $\bm{Z}$ \textbf{Parallel}}
\State Assign the workload of $\bm{Z}[j]$ to $\text{SCP}[j\%p_{sys}]$
\State load $\bm{Z}[j]$ to the Sparse Data Queue from Results Buffer
\For{each $e(i,j,value)$ in $\bm{X}[j]$ }  {\color{blue}\Comment{Scatter Phase}}
\State Fetch $\bm{Y}[i]$ from BufferO   {\color{blue}\Comment{ISN routes $e$ to BufferO}}
\State Form input pair ($\bm{Y}[i]$, $e$)  {\color{blue}\Comment{DSN routes input to SCPs}}
\EndFor

\For{each input pair ($\bm{Y}[i]$, $e$)}
{\color{blue}\Comment{Gather Phase}}
    \For{each non-zero $\bm{Y}[i][k]$ in $\bm{Y}[i]$}  {\color{blue}\Comment{SCP}}
    \State Produce $u \gets $ {Update($e.value \times \bm{Y}[i][k]$)}
    \State {Merge $\bm{Z}[j][k] \gets$ {Reduce($u$)}} 
    \EndFor
    
    \EndFor
\State Store $\bm{Z}[j]$ to the Result Buffer  {\color{blue}\Comment{Obtain $\bm{Z}[j]$}}
\EndFor
\end{algorithmic}
\end{small}
\end{algorithm}

\vspace{0.1cm}
\noindent \textbf{SPMM Mode}: The ALU array is organized as $p_{sys}$ parallel Sparse Computation Pipelines (SCP) as shown in Figure \ref{fig:execution-modes}. Each SCP has two ALUs to perform  multiplication of two non-zero elements and  the merging of intermediate results. Each SCP also has a Sparse Data Queue (SQ) to store the intermediate results in sparse format. The multiplication of two input sparse matrices is executed using the Row-wise Product  with Scatter-Gather paradigm as shown in Algorithm \ref{alg:scatter-gather-SPMM}. 
For Row-wise Product,  an row $ \bm{Z}[j]$  of  output matrix $\bm{Z}$ is calculated through: 
\begin{equation}
    \bm{Z}[j] = \sum_{i}\bm{X}[j][i]*\bm{Y}[i]
    \label{eq:workload-of-an-row}
\end{equation}
For calculating the output matrix $\bm{Z}$, a SCP is assigned the workload of an row of output matrix (Equation \ref{eq:workload-of-an-row}). $p_{sys}$ SCPs can calculate $p_{sys}$ output rows in parallel until all the rows of the output matrices are calculated.
To efficiently execute Row-wise Product, all input sparse matrices ($\bm{X}$, $\bm{Y}$) and output matrix are stored using COO format in row-major order (See Section \ref{subsec:data-format}). Using SPMM Mode, we can skip the zero elements in both the input matrices. SPMM Mode can execute $p_{sys}$ multiply-accumulate (MAC) operations per clock cycle. 

\vspace{0.1cm}
\noindent \textbf{Mode switching}: The execution mode is set by the control bits of the hardware multiplexers in ACM. The overhead of switching execution modes is just one clock cycle.

\vspace{0.1cm}
\noindent \textbf{Trade-off}: The three execution modes have different ways of dealing with  non-zero elements in the two input matrices (Section \ref{subsec:problem-define}). Therefore, their execution time of multiplying two input matrices depends on the data sparsity. We analyze the trade-off of the three execution modes w.r.t. data sparsity in Section \ref{subsec:performance-model}.


\subsubsection{Auxiliary Hardware Module (AHM)}
\label{subsubsec:AHM}
While the ACM can execute various primitives, the data format and layout should meet the requirement of the execution modes (Table \ref{tab:my_label}). Moreover, the soft processor  needs the data sparsity information at runtime for dynamic K2P mapping. To this end, the AHM has the following hardware modules:
(1) a Layout Transformation Unit and  a Layout Merger to transform the data layout, (2) a Sparsity Profiler (SP) to obtain the density of the intermediate results, (3) Format Transformation Module (FTM), which contains a Sparse-to-Dense  Module and a Dense-to-Sparse Module. 

\vspace{0.1cm}
\noindent \textbf{Layout Transformation Unit (LTU)}: Transformation of the data layout between row-major order and column-major order is transposing a matrix.  LTU is implemented using a streaming permutation network \cite{chen2015energy} (See \cite{chen2015energy}  for details) for efficient layout transformation.  Since most of the on-chip data are stored using row-major order, we store all the data partitions of ($\bm{A}$, $\bm{H}$, $\bm{W}$) in the external memory using row-major order to minimize the effort for data layout transformation. 

\vspace{0.1cm}
\noindent \textbf{Layout Merger}: When the accelerator executes a task  (See algorithm \ref{alg:Computation-Task}), the results  $\bm{Z}$ can be in row-major or column-major order. Therefore, in Results Buffer, we store two partial results of $\bm{Z}$ in row-major and column-major order, respectively. The two partial results of  $\bm{Z}$ are merged by Layout Merger into row-major order when $\bm{Z}$ is sent back to the external memory.
Note that the LTU is also used by BufferO to transform the data layout for $\bm{X}_{2}^{\intercal}$ (column-major order of $\bm{X}_{2}$).


\vspace{0.1cm}
\noindent \textbf{Sparsity Profiler}: To profile the density of sparse matrix or dense matrix, we use the adder tree based design for the Sparsity Profiler. At the output port of the Result Buffer, we implement a comparator array with an adder tree to count the total number of non-zero elements. After obtaining the data sparsity of the current output matrix, the sparsity information is sent to the soft processor.

\begin{figure}[h]
     \centering
     \includegraphics[width=8.5cm]{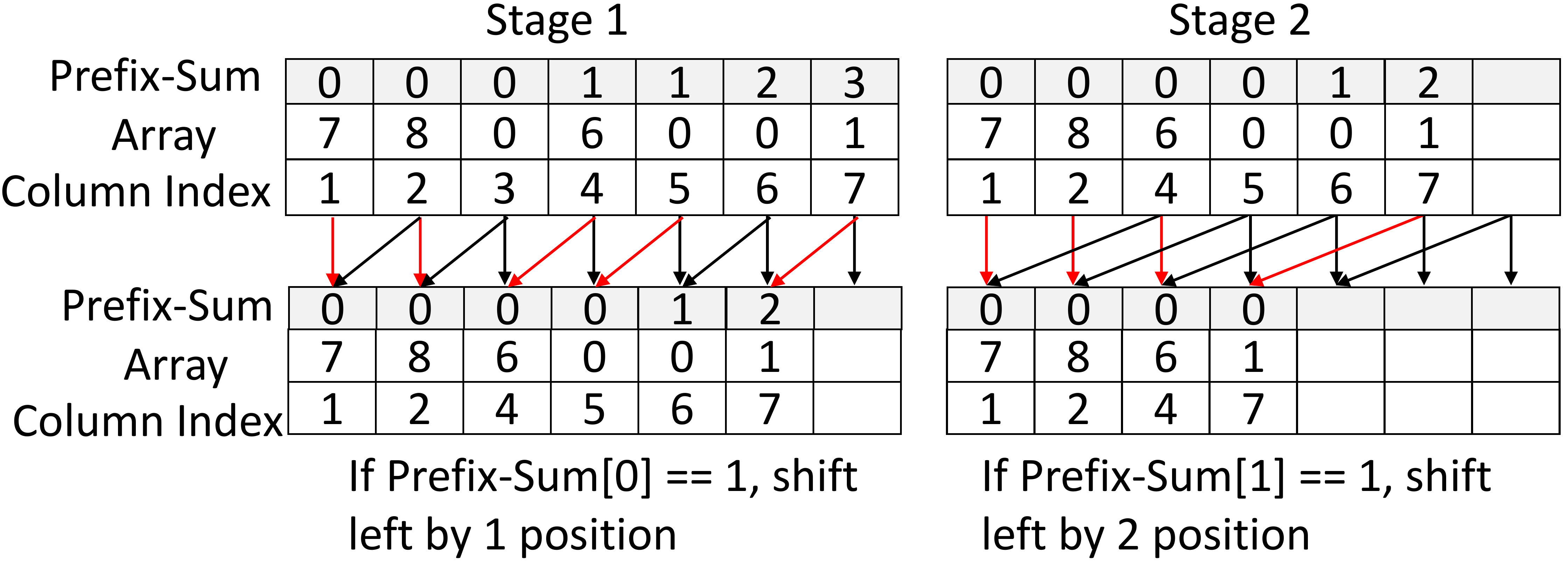}
     \caption{Transforming dense format to sparse format}
     \label{fig:dense-to-sparse}
\end{figure}

\vspace{0.1cm}
\noindent \textbf{Dense-to-Sparse (D2S) Module}: It transforms an array from dense format to sparse format. Suppose the D2S Module can read $n$ elements per clock cycle. Then, the D2S Module has $\log(n)$ pipeline stages.   For an $n$-element array, we use the value of Prefix-Sum to indicate the number of zeros before an element in this array. An example is shown in Figure \ref{fig:dense-to-sparse}. In Stage $i~(1 \leqslant i \leqslant \log(n))$, an array element will be shifted left by $2^{i-1}$ positions if the $(i-1)^{\text{th}}$ bit of Prefix Sum value is equal to 1. The throughput of D2S Module is $n$ elements per cycle. For example, a DDR4 channel of the FPGA board can output 16 32-bit data per cycle. A D2S Module of $n = 16$ is sufficient to match the data rate of a DDR4 channel.  
The architecture of S2D is similar to D2S, but in the reverse direction. 

\vspace{0.2cm}
{
\subsubsection{Double Buffering} \label{subsubsec-double-buffering} We exploit double buffering technique for Buffer[U/O/P] and Results Buffer. Therefore, when the Computation Core is executing the current task, the Buffers can load the input data of the next task. The data sparsity profiling, data layout and format transformation are streaming processes that can be executed during the data loading/storing process. Double Buffering not only overlaps the computation and data communication, but also hides the overhead of sparsity profiling and data layout/format transformation.
}

%% file: 6-runtime-system.tex
\section{Runtime system}
\label{sec:runtime-system}

\subsection{Performance Model}
\label{subsec:performance-model}

The performance model predicts the execution time of the primitives for a given data sparsity. For analysis, we denote the two input matrices to a Computation Core as $\bm{X}\in \mathbb{R}^{m\times n}$ and $\bm{Y} \in \mathbb{R}^{n\times d}$ where $\bm{X}$ has the density  $\alpha_{\bm{X}}~(0 \leqslant \alpha_{\bm{X}} \leqslant 1)$  and $\bm{Y}$ has the density  $\alpha_{\bm{Y}}~(0 \leqslant \alpha_{\bm{Y}} \leqslant 1)$. 

\begin{table}[h]
\centering
\caption{Performance model}
\begin{adjustbox}{max width=0.48\textwidth}
\begin{tabular}{cccc}
\toprule
 & \textbf{GEMM} & \textbf{SpDMM} & \textbf{SPMM}\\
 \midrule \midrule
MACs per cycle & $p_{sys}^{2}$  & $p_{sys}^{2}/2$ &  $p_{sys}$  \\ \midrule
 \begin{tabular}[|c|]{@{}c@{}}  Execution time  \\ (cycles)\end{tabular} & $\frac{mnd}{p_{sys}^{2}}$  & \begin{tabular}[|c|]{@{}c@{}}  $\alpha_{\text{min}} \frac{2mnd}{p_{sys}^{2}}$, where \\ $\alpha_{\text{min}} = \text{Min}(\alpha_{\bm{X}}, \alpha_{\bm{Y}})$\end{tabular}  &  $\alpha_{\bm{X}}\alpha_{\bm{Y}} \frac{mnd}{p_{sys}}$  \\ \bottomrule
\end{tabular}
\end{adjustbox}
\end{table}

In the GEMM mode, the two input matrices are viewed as dense matrices and the Computation Core can execute $p_{sys}^{2}$ MACs per cycle. Therefore, the total execution time is $\frac{mnd}{p_{sys}^{2}}$ cycles.  In the SpDMM mode, the Computation Core can skip the zero elements in one input matrix and can execute $p_{sys}^{2}/2$ MACs per cycle. We view the input matrix with lower density as a sparse matrix and view another input matrix as a dense matrix. Therefore, the total execution time is $\alpha_{\text{min}} \frac{2mnd}{p_{sys}^{2}}$ cycles where $\alpha_{\text{min}} = \text{Min}(\alpha_{\bm{X}}, \alpha_{\bm{Y}})$.
In the SPMM mode, the Computation Core can skip the zero elements in both two input matrices and can execute $p_{sys}$ MACs per cycle. Therefore, the total execution time is $\alpha_{\bm{X}}\alpha_{\bm{Y}} \frac{mnd}{p_{sys}}$ cycles. In the state-of-the-art FPGA such as Xilinx Alveo U250, the dimension of a Computation Core $p_{sys}$ can be chosen to be $\geqslant 8$. We denote $\alpha_{max} = \text{Max}(\alpha_{\bm{X}}, \alpha_{\bm{Y}})$.
To summarize, for executing $\bm{Z} = \bm{X}\times \bm{Y}$ on a Computation Core, when $\alpha_{\text{min}} \geqslant \frac{1}{2}$, GEMM Mode has the least execution time; When $\alpha_{\text{min}} < \frac{1}{2}$ and $\alpha_{max} \geqslant \frac{2}{p_{sys}}$, SpDMM Mode has the least execution time; When $\alpha_{\text{min}} < \frac{1}{2}$ and $\alpha_{max} < \frac{2}{p_{sys}}$, SPMM Mode has the least execution time. The three cases are non-overlapping and cover all the points in the domain $0\leqslant \alpha_{min} \leqslant \alpha_{max} \leqslant 1$.

\subsection{Dynamic Kernel-to-primitive Mapping}
\label{subsec:dyna-k2p-mapping}

\begin{algorithm}
\caption{Dynamic kernel-to-primitive (K2P) mapping Algorithm for a computation task}
\label{alg:Kernal-to-mapping-Algorithm}
\begin{algorithmic}[1]
 \renewcommand{\algorithmicrequire}{\textbf{Input:}}
\renewcommand{\algorithmicensure}{\textbf{Output:}}
 \Require $\{ \bm{X}_{i1}$, $\bm{X}_{i2}$, ..., $\bm{X}_{iK} \}$ and $\{\bm{Y}_{1j}$, $\bm{Y}_{2j}$, ..., $\bm{X}_{Kj} \}$;
\For{$t=1$ to $K$}
    \State{TargetPrimitive($\bm{X}_{it}$, $\bm{Y}_{tj}$) $\leftarrow$ NULL}
    \State{The buffers to store $\bm{X}_{it}$ and $\bm{Y}_{tj}$: $B_{\bm{X}_{it}}$, $B_{\bm{Y}_{tj}}$}
    \State{$\alpha_{\text{min}} = \text{Min}(\alpha_{\bm{X}_{it}}, \alpha_{\bm{Y}_{tj}})$}   {\color{blue}\Comment{$\alpha_{\bm{X}_{it}}$: The density of $\bm{X}_{it}$}}
    \State{$\alpha_{\text{max}} = \text{Max}(\alpha_{\bm{X}_{it}}, \alpha_{\bm{Y}_{tj}})$}  {\color{blue}\Comment{$\alpha_{\bm{Y}_{tj}}$: The density of $\bm{Y}_{tj}$}}
    \If{$\alpha_{\text{min}} = 0$}   {\color{blue}\Comment{Skip empty input matrix}}
        \State{Skip the multiplication of $\bm{X}_{it}$ and $\bm{Y}_{tj}$}
    \EndIf
    \If{$\alpha_{\text{min}} \geqslant  \frac{1}{2}$}
        \State{TargetPrimitive($\bm{X}_{it}$, $\bm{Y}_{tj}$)   $\leftarrow$ GEMM}
        \State{ $B_{\bm{X}_{it}}$ $\leftarrow$ BufferO and   $B_{\bm{Y}_{tj}}$ $\leftarrow$ BufferP}
    \Else
        \If{$\alpha_{\text{max}} \geqslant \frac{2}{p_{sys}}$}
            \State{TargetPrimitive($\bm{X}_{it}$, $\bm{Y}_{tj}$)  $\leftarrow$ SpDMM}
            \State{$B_{\text{argmin}(\alpha_{M})}$ $\leftarrow$ BufferU, ($M\in \{\bm{X}_{it}, \bm{Y}_{tj}\}$)}
            \State{$B_{\text{argmax}(\alpha_{M})}$ $\leftarrow$ BufferO, ($M\in \{\bm{X}_{it}, \bm{Y}_{tj}\}$)}
        \Else
            \State{TargetPrimitive($\bm{X}_{it}$, $\bm{Y}_{tj}$)  $\leftarrow$ SPMM}
            \State{$B_{\bm{X}_{it}}$ $\leftarrow$ BufferU and $B_{\bm{Y}_{tj}}$ $\leftarrow$ BufferO}
        \EndIf
    \EndIf
\EndFor
\end{algorithmic}
\end{algorithm}

The Analyzer performs dynamic kernel-to-primitive (K2P) mapping for each computation task shown in Algorithm \ref{alg:Kernal-to-mapping-Algorithm}. For each pair of input matrices ($\bm{X}_{it}$, $\bm{Y}_{tj}$), the runtime system fetches their densities $\alpha_{\bm{X}_{it}}$ and $\alpha_{\bm{Y}_{tj}}$. Then, the Analyzer determines the target  primitive for multiplying $\bm{X}_{it}$ and $\bm{Y}_{tj}$, and also determines which buffers to store $\bm{X}_{it}$ and $\bm{Y}_{tj}$. The proposed dynamic K2P algorithm has the computation complexity $\mathcal{O}(K) = \mathcal{O}(\frac{|\mathcal{V}|}{N_{1}} + \frac{f_{1}}{N_{2}})$ for a computation task, which has small overhead compared with total computation complexity of a task $\mathcal{O}(|\mathcal{V}|*N_{2} + f_{1}*N_{2}^{2})$. See evaluation results in Section \ref{subsec:analysis-compiler-runtime}.
There are several benefits: (1) the proposed dynamic K2P mapping is fine-grained that for different data partitions, we can use different primitives to efficiently exploit the data sparsity in the input. (2) When the accelerator is executing kernel $l$, the runtime system can perform K2P mapping for kernel $l+1$. Therefore, the overhead of the runtime system can be hidden.


\subsection{Task Scheduling}
\label{subsec:task-scheduling}

The scheduler performs scheduling of computation tasks (See Section \ref{sucsec:IR}) on the parallel Computation Cores as shown in Algorithm \ref{alg:scheduling-Scheduler}. The proposed task scheduling is a \emph{dynamic task scheduling strategy}. Each Computation Core maintains an interrupt interface to trigger the interrupt handling in the soft processor when the Computation Core is idle. Then, the soft processor assigns a task to the Computation Core. { { \color{red}}

\begin{algorithm}
\caption{Task scheduling}
\label{alg:scheduling-Scheduler}
\begin{algorithmic}[1]
 \renewcommand{\algorithmicrequire}{\textbf{Input:}}
\renewcommand{\algorithmicensure}{\textbf{Output:}}
 \Require Intermediate Representation of the GNN model: IR; The number of computation kernels in the IR: $L$;
 \Ensure Output of the GNN model;
\For{$l=1$ to $L$}
    \For{each \emph{Task} in kernel $l$ of IR \textbf{parallel}}
        \If{there is an idle CC: $\text{CC}_{i}$}
            \State Assign this \emph{Task} to $\text{CC}_{i}$
            \State $\text{CC}_{i}$ executes this computation \emph{Task}
        \EndIf
    \EndFor 
    \State Wait until all the \emph{Tasks} in kernel $l$ are executed 
\EndFor
\end{algorithmic}
\end{algorithm}
 
\noindent { \textbf{Partition size ($N_{1}$, $N_{2}$)}:  The objectives of the data partitioning are to (1) enable fine-grained data sparsity exploitation, (2) exploit data locality, and (3) maximize resource utilization during dynamic task scheduling (Algorithm \ref{alg:scheduling-Scheduler}). 
Specifically, to maximize resource utilization that keeps all the Computation Cores busy, the compiler selects the partition configuration ($N_{1}, N_{2}$) such that there will be at least $\eta*N_{CC}$ ($\eta \geq 1$) tasks in each computation kernel assigned to $N_{CC}$ Computation Cores. $\eta$ is a factor that is determined empirically. Since different partitions can have different data sparsity leading to the different workloads of the tasks, small $\eta$ (e.g., $\eta = 1$) can potentially lead to long idle time for the Computation Cores with small workloads. Therefore, we set  $\eta = 4$ following state-of-the-art graph processing frameworks \cite{lakhotia2020gpop}.

To meet the above three objectives, we use a heuristic approach to determine the partition size as shown in Algorithm \ref{alg:partition-size}. As shown in Algorithm \ref{alg:scheme-aggregate} (line 2-3), the number of tasks of an Aggregate kernel is $\mathcal{T}_{a} = \frac{|\mathcal{V}|*f_{1}}{N_{1}*N_{2}}$. Also, as shown in Algorithm \ref{alg:scheme-Update} (line 2-3), the number of tasks of an Update kernel is $\mathcal{T}_{u} =  \frac{|\mathcal{V}|*f_{2}}{N_{2}*N_{2}}$. For simplicity, we use 
 $Q$ to denote the workload of a kernel (e.g., $Q=|\mathcal{V}|*f_{1}$ or $Q=|\mathcal{V}|*f_{2}$), and use $Q[k]$ to denote the workload of $k^{\text{th}}$ ($1 \leqslant k \leqslant L$) kernel. We use $p()$ to denote the function that determines the number of tasks of a kernel based on $Q$, $N_{1}$, and $N_{2}$. For example, $\mathcal{T}_{a} = p(Q, N_{1}, N_{2}) = \frac{Q}{N_{1}*N_{2}}$ and $\mathcal{T}_{u} = p(Q, N_{2}) = \frac{Q}{N_{2} * N_{2} }$. In line 9 and line 15 of Algorithm \ref{alg:partition-size}, the partition size of each kernel is constrained by $N_{it} = \text{min}(N_{it}, N_{\text{max}})$, where $\text{min}(N', N_{\text{max}})$ is the largest partition size such that $N_{it} \leqslant N'$ and $N_{it}  \leqslant N_{\text{max}}$. $N \leqslant N'$ ensures that there will be at least $\eta*N_{CC}$ tasks of a kernel for load balance. $N_{it}  \leqslant N_{\text{max}}$ ensures that the data partition does not exceed the size of on-chip memory. Lines 10 and 16  find a partition size $N_{1}$ and $N_{2}$ that can be used for all the kernels.

}

\begin{algorithm}
\caption{Data partitioning algorithm}
\label{alg:partition-size}
\begin{small}
\begin{algorithmic}[1]
 \renewcommand{\algorithmicrequire}{\textbf{Input:}}
\renewcommand{\algorithmicensure}{\textbf{Output:}}
\Require On-chip memory size $S_{o}$; Computation workload of each kernel: $\{Q[k]: 1 \leqslant k \leqslant L \}$; $p()$: function that determines the number of tasks of a kernel based on $Q$, $N_{1}$ and $N_{2}$; $g()$:  function that determines the maximum partition size based on the on-chip memory size $S_{o}$; $\eta$: factor for load balance.
\Ensure Partition size $N_{1}$, $N_{2}$; 
 \State $N_{\text{max}} \leftarrow  g(S_{o})$    {\color{blue}\Comment{Maximum partition size}}
  \State {\color{blue} //Objective: Maximize $N_{1}$ and $N_{2}$ to improve data locality}
 \State {\color{blue}//Constraint 1 (Maximize utilization): $\mathcal{T}_{a}$, $\mathcal{T}_{u}  \geq \eta *N_{CC}$}
 \State {\color{blue}//Constraint 2 (Memory capacity): $N_{1}$, $N_{2}  \leqslant N_{\text{max}}$ }
 \State {\color{brown} ======= Step 1: determine $N_{2}$ ========} 
 \State $N_{2} \leftarrow N_{\text{max}}$ 
 \For{each Update kernel: $k^{\text{th}}$ kernel}
     \State Choose largest $N'$ such that $\mathcal{T}_{u}[k] = p(Q[k], N') = \eta *N_{CC}$
     \State $N_{it} \leftarrow \text{min}(N', N_{\text{max}})$  
     \State $N_{2} \leftarrow \text{min}(N_{it}, N_{2})$
 \EndFor
  \State {\color{brown} ======= Step 2: determine $N_{1}$ ========} 
 \State $N_{1} \leftarrow N_{\text{max}}$ 
 \For{each Aggregate kernel: $k^{\text{th}}$ kernel}
     \State Choose largest $N'$ such that $\mathcal{T}_{a}[k] = p(Q[k], N', N_{2}) =$  $~~~~~\eta *N_{CC}$
     \State $N_{it} \leftarrow \text{min}(N', N_{\text{max}})$  
     \State $N_{1} \leftarrow \text{min}(N_{it}, N_{1})$
 \EndFor
\end{algorithmic}
\end{small}
\end{algorithm}

%% file: 7-analaysis.tex

\section{Implementation details}
\label{sec:Implementation details}

\begin{figure}[h]
     \centering
     \includegraphics[width=8.7cm]{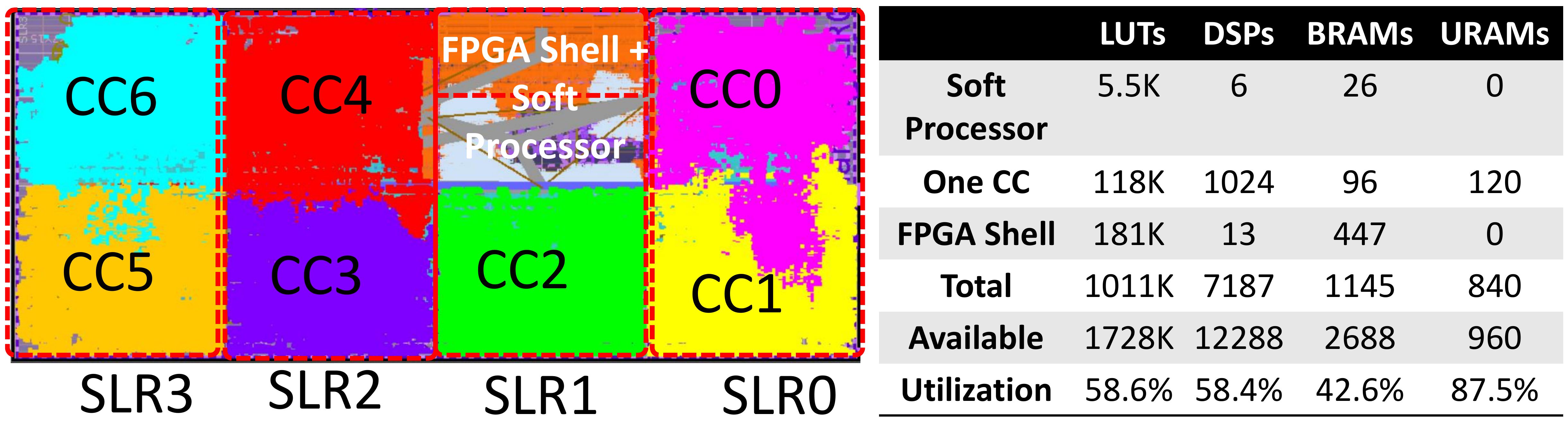}
     \caption{The layout (FPGA chip) and resource utilization of the proposed design on Xilinx Alveo U250. The Computation Cores (CC0-CC6) are represented using different colors.}
     \label{fig:schematic}
\end{figure}

We implement the proposed accelerator on a state-of-the-art FPGA board -- Xilinx Alveo U250, which
has four Super Logic Regions (SLR) \cite{ref-alvelu250}.  As shown in Figure \ref{fig:schematic}, we implement two Computation Cores (CC) in each SLR except for SLR1, because the FPGA shell (which handles the CPU-FPGA communication) and soft processor is placed in SLR1. For each CC, $p_{sys}=16$. We develop the CC using Verilog HDL, and implement the soft processor using Xilinx Microblaze Soft IP core \cite{Microblaze-link}. Each CC is connected to the soft processor through the AXI4-Stream interface \cite{Microblaze-link}, through which the soft processor sends the control signals to CC and the CC sends the sparsity information to the soft processor. 
{ We develop the compiler using Python. The IR of a kernel is implemented as a Python object that stores the meta data of a kernel and its execution scheme.}
We develop the Runtime system on the soft processor using C in Xilinx Vitis Unified Software Platform (version 2020.1). The Index Shuffle Network and Data Shuffle Network are implemented using a butterfly network with buffering to handle the routing congestion. We perform synthesis and Place\&Route using Vivado 2020.1.  The resource utilization is shown in Figure \ref{fig:schematic}. The CCs run at 250 MHz.

\noindent \textbf{Soft processor}: Our implementation achieves 370 MHz and  around 500 Million Instructions Per Second \cite{Microblaze-link} performance. It has two caches -- an Instruction Cache (I-Cache) and a Data Cache (D-Cache). I-Cache has size 32 KB which is sufficient to hold the binary code of the runtime system after a warm-up execution. D-Cache has the size 64 KB which stores the sparsity of the data partitions. For large graphs that D-Cache is not enough to hold the sparsity information of all the data partitions, we store it in the external memory and prefetch the sparsity information to the D-Cache. The soft processor reads/writes the data from/to the AXI-stream interface through the $\verb|get|$ and $\verb|put|$ instructions \cite{Microblaze-link}, which have one or two clock cycles latency.


%% file: 8-experiment.tex
\section{Evaluation Results}
\label{sec:Evaluation-Results}

 {This Section is organized as follows: In Section \ref{subsec:Impact-of-K2P-mapping}, we measure of the impact of the dynamic K2P mapping strategy. In Section \ref{subsec:analysis-compiler-runtime}, we analyze the overhead of compilation and runtime system. In Section \ref{subsec:comparison-with-SOTA}, we compare our work with the state-of-the-art implementations.}

\subsection{Benchmarks and Baselines}

\begin{figure}[h]
     \centering
     \includegraphics[width=8.5cm]{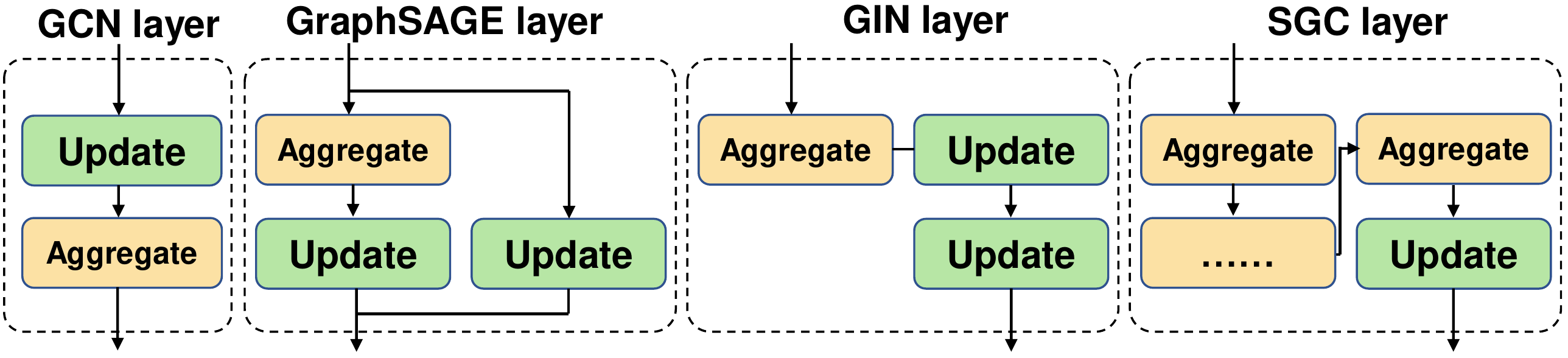}
     \caption{The IR of various GNN layers}
     \label{fig:IR-GNN-layer}
\end{figure}

\noindent \textbf{Benchmarks}: We evaluate Dynasparse on four widely used GNN models -- GCN \cite{kipf2016semi}, GraphSAGE (SAGE) \cite{hamilton2017inductive}, GIN \cite{xu2018powerful}, and SGC \cite{wu2019simplifying}. 
Figure \ref{fig:IR-GNN-layer} shows the IR of various GNN layers. 
We evaluate the design on six widely used graph datasets -- Cora (CO) \cite{kipf2016semi}, CiteSeer (CI) \cite{kipf2016semi}, PubMed (PU) \cite{kipf2016semi}, Flickr (FL) \cite{zeng2019graphsaint}, NELL (NE) \cite{yang2016revisiting}, Reddit (RE) \cite{hamilton2017inductive}. We evaluate the 2-layer GNN models used in \cite{kipf2016semi, geng2020awb, yan2020hygcn, zhang2021boostgcn}, where the hidden dimension for CO, CI and PU is set as 16, and the hidden dimension for FL, NE and RE is set as 128.

\vspace{0.1cm}
\noindent \textbf{Baselines}: We compare our work with the state-of-the-art CPU (AMD Ryzen 3990x), GPU (Nvidia RTX3090) and GNN accelerators  HyGCN \cite{yan2020hygcn}, BoostGCN \cite{zhang2021boostgcn}. The details of the platforms are shown in Table \ref{tab:platform-specifications}.



\begin{table}[!ht]
\centering
\caption{Specifications of platforms }
\begin{threeparttable}
\begin{adjustbox}{max width=0.48\textwidth}
\begin{tabular}{c|ccccc}
 \toprule
& CPU & GPU  & \cite{yan2020hygcn} & \cite{zhang2021boostgcn}  & Dynasparse  \\  \midrule
\midrule 
Platform  &  \begin{tabular}[|c|]{@{}c@{}} Ryzen \\ 3990x\end{tabular} & \begin{tabular}[|c|]{@{}c@{}} Nvidia\\ RTX3090 \end{tabular} & ASIC   & \begin{tabular}[|c|]{@{}c@{}}  Stratix 10 \\ GX\end{tabular}   & \begin{tabular}[|c|]{@{}c@{}} Alveo \\ U250 \end{tabular}  \\  
 \rowcolor{LightCyan} {Technology}  & \begin{tabular}[|c|]{@{}c@{}} TSMC \\ 7 nm  \end{tabular}  & \begin{tabular}[|c|]{@{}c@{}} TSMC \\ 7nm  \end{tabular} & \begin{tabular}[|c|]{@{}c@{}} TSMC \\12 nm  \end{tabular}& \begin{tabular}[|c|]{@{}c@{}} Intel \\ 14 nm \end{tabular} & \begin{tabular}[|c|]{@{}c@{}}  TSMC \\ 16 nm \end{tabular}\\ 
{Frequency} & 2.90 GHz  & 1.7 GHz & 1 GHz  & 250 MHz & 250 MHz \\  
 \rowcolor{LightCyan}
 \begin{tabular}[|c|]{@{}c@{}}  Peak Performance \\ (TFLOPS)
 \end{tabular}& 3.7 & 36  &  4.608   & 0.64 & 0.512  \\ 
{On-chip Memory}& 256 MB & 6 MB & 35.8 MB&  32 MB & 45 MB   \\ 
 \rowcolor{LightCyan}
{Memory Bandwidth}& 107 GB/s & 936.2 GB/s  & 256 GB/s &  77 GB/s  & 77 GB/s \\ \bottomrule
\end{tabular}
\end{adjustbox}
\end{threeparttable}
\label{tab:platform-specifications}
\end{table}

\begin{table}[ht]
\centering
\caption{Dataset Statistics}
\begin{adjustbox}{max width=0.49\textwidth}
\begin{tabular}{ccccccc}
\toprule
\textbf{Dataset}   & \textbf{Vertices} & \textbf{Edges} & \textbf{Features} & \textbf{Classes} &  \begin{tabular}[|c|]{@{}c@{}}  \textbf{Density of} \\ $\bm{A}$ \end{tabular} & \begin{tabular}[|c|]{@{}c@{}}  \textbf{Density of} \\ $\bm{H}^{0}$ \end{tabular} \\
\midrule
\midrule
 CI& 3327 & 4732& 3703& 6 &  $0.08\%$ & $0.85\%$\\
 \rowcolor{LightCyan}
 CO & 2708 &  5429& 1433 & 7 & $0.14\%$ & $1.27\%$ \\
 PU& 19717 &  44338 & 500 & 3 & $0.02\%$ & $10.0\%$ \\
 \rowcolor{LightCyan}
 FL & 89,250 &    899,756 & 500 & 7 & $0.01\%$ & $46.4\%$ \\
 NE& 65,755 &  251,550 & 61,278 & 186 & $0.0058\%$ & $0.01\%$\\
 \rowcolor{LightCyan}
 RE& 232,965  & $11\times 10^{7}$ & 602 & 41 & $0.21\%$ & $100.0\%$ \\
\bottomrule
\end{tabular}
\end{adjustbox}
\label{tab:datasets-statistics}
\end{table}

\noindent \textbf{Performance metric}: Following the convention in \cite{geng2020awb, yan2020hygcn, zhang2021boostgcn}, we use \emph{latency} ({accelerator execution latency}) as the  metric which is the duration from the time when the accelerator starts to execute the optimized IR to the time all the inference results are obtained. The preprocessing time by the compiler is not included in the latency, because (1) the overhead of generating the optimized IR is usually small (See Section \ref{subsec:analysis-compiler-runtime}) and the optimized IR can be stored and reused if the sparsity of the input graph and GNN model changes, (2)  we follow the same convention in \cite{geng2020awb, yan2020hygcn, zhang2021boostgcn} for a fair comparison.

\begin{table}[]
\centering
\caption{The latency (ms) on the unpruned GNN models}
\begin{adjustbox}{max width=0.48\textwidth}
\begin{tabular}{llllllll}
\toprule
                  &  & CI & CO & PU & FL & NE & RE \\ \midrule \midrule
\multirow{3}{*}{GCN \cite{kipf2016semi}} & 
        \texttt{S1} & 31E-1& 9.6E-1&2.7E-1& 10E0&83E2&9.3E1\\
        & \texttt{S2} & 8.9E-3 & 5.6E-3 & 7.1E-3 & 9.9E0 & 5.4E0 & 12E1\\
        & \texttt{Dynamic} & 7.7E-3& 4.7E-3&6.3E-2& 8.8E0&2.9E0&8.4E1\\ \cmidrule{2-8}
        & SO-S1 &$41.3\times$&$21.5\times$&$4.29\times$&$1.13\times$&$278\times$&   $1.10\times$ \\
        & SO-S2 &$1.15\times$&$1.19\times$&$1.12\times$&$1.11\times$&$1.82\times$&   $1.42\times$ \\  \midrule
\multirow{3}{*}{SAGE \cite{hamilton2017inductive}} & \texttt{S1}       
        &74E-2&25E-2&65E-2&20E0&17E2&334E0\\
       & \texttt{S2} &75E-2&25E-2&69E-2&28E0&17E2&389E0\\
        &  \texttt{Dynamic}&33E-2&11E-2&42E-2&19E0&83E1&331E0\\\cmidrule{2-8}
        &  SO-S1 & $1.93\times$ & $1.72\times$ & $1.56\times$ & $1.02\times$ & $2.05\times$ & $1.01\times$ \\ 
        &  SO-S2 & $1.94\times$ & $1.73\times$ & $1.65\times$ & $1.41\times$ & $2.05\times$ & $1.17\times$ \\ \midrule
\multirow{3}{*}{GIN \cite{xu2018powerful}} &
        \texttt{S1} &4.3E-1&1.5E-1&4.1E-1&1.3E1&8.8E2&3.1E2 \\
      & \texttt{S2} &7.4E-1&2.4E-1&6.5E-1&2.0E1&1.7E3&3.4E2\\
      & \texttt{Dynamic}  & 3.3E-1 & 1.1E-1 & 3.7E-1 & 1.2E1 & 8.3E2 & 2.7E2 \\\cmidrule{2-8}
        & SO-S1  & $1.30 \times$ & $1.40\times$  & $1.11\times$  & $1.13\times$ & $1.06\times$ & $1.15\times$   \\
        & SO-S2 & $2.26 \times$ & $2.31\times$  & $1.76\times$  & $1.73\times$ & $2.05\times$ & $1.25\times$   \\ \midrule
\multirow{3}{*}{SGC \cite{wu2019simplifying}} 
        & \texttt{S1}  & 5.3E-1 & 2.0E-1 & 5.5E-1 & 1.29E-1 & 9.33E2 & 5.7E2 \\
        & \texttt{S2}  & 8.5E-1 & 3.0E-1 & 7.9E-1 & 2.18E-1 & 1.77E3 & 6.0E2 \\
        & \texttt{Dynamic}   & 4.3E-1 & 1.5E-1 & 5.1E-1 & 1.27E-1 & 8.83E2 & 5.0E2 \\\cmidrule{2-8}
        & SO-S1  & $1.23 \times$ & $1.27 \times$ &  $1.08 \times$ & $1.02 \times$ & $1.06 \times$ & $1.13 \times$   \\
        & SO-S2 & $1.95 \times$ & $1.91 \times$ &  $1.55 \times$ & $1.72 \times$ & $1.99 \times$ & $1.19 \times$   \\
                  \bottomrule
\end{tabular}
\end{adjustbox}
\label{tab:results-unpruned-GNN-model}
\end{table}

\subsection{Impact of Dynamic K2P Mapping Strategy}
\label{subsec:Impact-of-K2P-mapping}

To demonstrate the impact of the proposed dynamic K2P mapping strategy, we execute the  following three K2P mapping strategies on our proposed accelerator: 
\begin{itemize}
    \item $\verb|Static-1|$ ($\verb|S1|$): It is used in \cite{yan2020hygcn, zhang2021boostgcn} that Aggregate() is mapped to SpMM  and Update() is mapped to GEMM.
    \item $\verb|Static-2|$ ($\verb|S2|$): It is used in \cite{geng2020awb} that both the Aggregate() and Update() are mapped to SpDMM. For Aggregate($\bm{A}$, $\bm{H}$), it views $\bm{A}$ as sparse matrix and views $\bm{H}$ as dense matrix. For Update($\bm{H}$, $\bm{W}$), it views $\bm{H}$ as sparse matrix and views $\bm{W}$ as dense matrix.
    \item $\verb|Dynamic|$:  It is our proposed dynamic K2P mapping strategy (Algorithm \ref{alg:Kernal-to-mapping-Algorithm}).
\end{itemize}

We use SO-S1 to denote the speedup of  \texttt{Dynamic} over \texttt{S1}. We use  SO-S2 to denote the speedup of  \texttt{Dynamic} over \texttt{S2}.

\vspace{0.1cm}
\noindent \textbf{Evaluation on unpruned GNN models}: We evaluate the above three strategies using unpruned GNN models where all the weight matrices have density $100\%$. The results are shown in Table \ref{tab:results-unpruned-GNN-model}. 
Compared with $\verb|S1|$ and $\verb|S2|$, $\verb|Dynamic|$ achieves $2.13 \times$ and $1.59 \times$ speedup on the average (geometric  mean), respectively.  $\verb|Dynamic|$ achieves limited speedup over $\verb|S2|$ on GCN because (1) for the first Update($\bm{H}^{0}$, $\bm{W}^{1}$) kernel of GCN, there is high data sparsity in $\bm{H}^{0}$ of CI, CO, PU and NE (See Table \ref{tab:datasets-statistics}), (2) both $\verb|Dynamic|$ and $\verb|S2|$ can exploit the sparsity of feature matrix $\bm{H}^{0}$ while $\verb|S1|$ does not exploit the sparsity of $\bm{H}^{0}$. As the first Update($\bm{H}^{0}$, $\bm{W}^{1}$) kernel of GCN consumes majority of the execution time,  $\verb|Dynamic|$ achieves very large speedup over $\verb|S1|$ on GCN. Since the weight matrices have density $100\%$, both $\verb|Dynamic|$ and $\verb|S2|$ map Update($\bm{H}^{0}$, $\bm{W}^{1}$) to SpDMM (for  CI, CO, PU and NE), leading to similar performance of $\verb|Dynamic|$ and $\verb|S2|$ on GCN.

\begin{figure}[h]
     \centering
     \includegraphics[width=8.5cm]{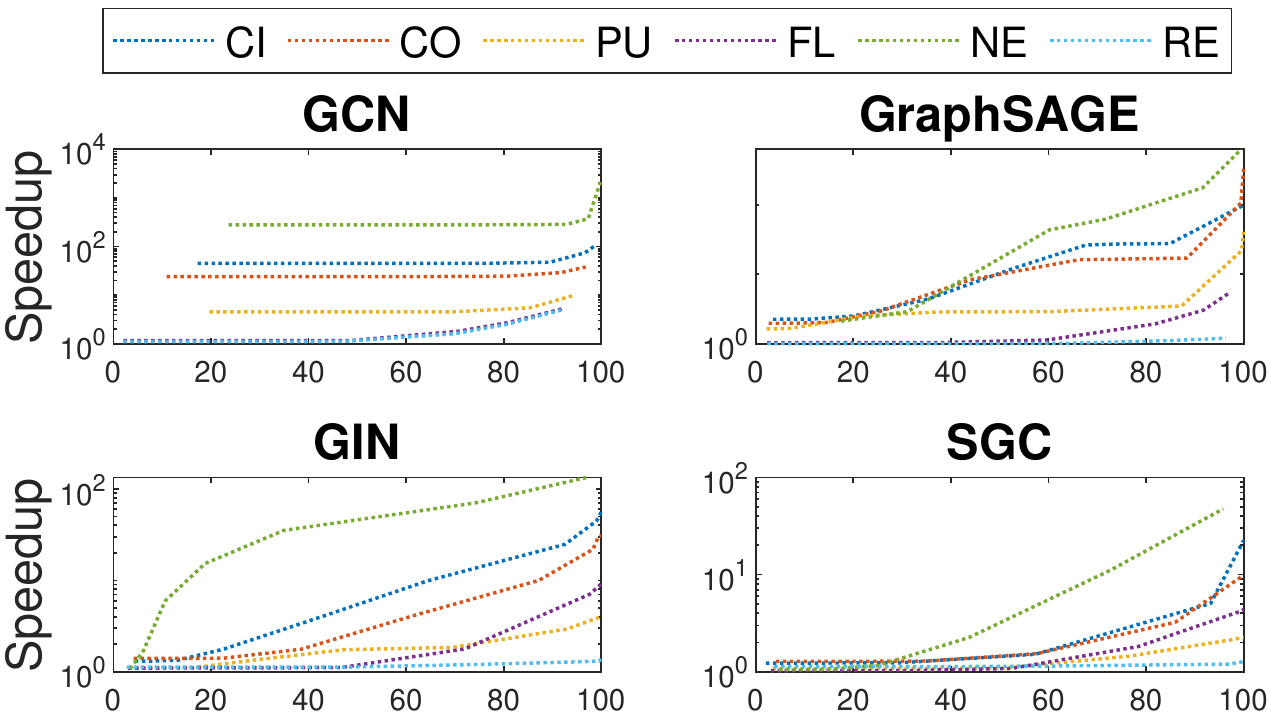} 
     \caption{Speedup of \texttt{Dynamic} over \texttt{S1} when there are various sparsity (\%) in the GNN weight matrices (X-axis)}
     \label{fig:model-sparse}
 \end{figure}
\begin{figure}[h]
     \centering
     \includegraphics[width=8.5cm]{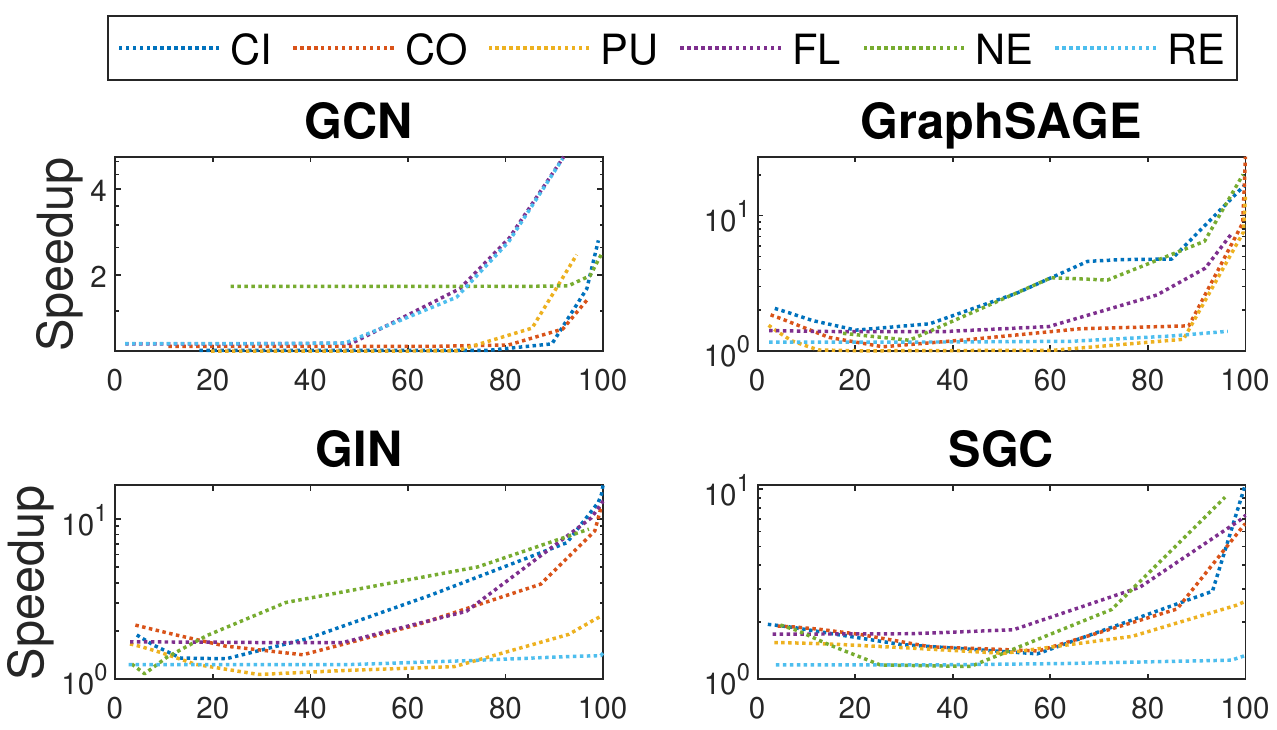} 
     \caption{Speedup of \texttt{Dynamic} over \texttt{S2} when there are various sparsity (\%) in the GNN weight matrices (X-axis)}
     \label{fig:speedup-over-S2}
 \end{figure}

\vspace{0.1cm}
\noindent \textbf{Evaluation on pruned GNN models}: We evaluate the three strategies using the pruned GNN models \cite{rahman2022triple} where the weight matrices are pruned to have various sparsity.  
Figures \ref{fig:model-sparse} and  \ref{fig:speedup-over-S2}   show the speedup of  \texttt{Dynamic} over \texttt{S1\&2}. For evaluation, all the weight matrices in a GNN model are pruned to have the same sparsity, and the sparsity of weights in Figures \ref{fig:model-sparse}\&\ref{fig:speedup-over-S2} means the average sparsity of all the weight matrices in a GNN model. Table \ref{tab:average-speedup} summarizes the average (geometric  mean) speedup under various sparsity of weight matrices. The achieved speedup over $\verb|S1|$ is because  $\verb|S1|$ cannot exploit the data sparsity in feature matrices and weight matrices. The achieved speedup over $\verb|S2|$ is due to (1) when there is limited data sparsity (density $<50\%$) in Update(), executing  Update() using SpDMM primitive is not efficient. (2) In Aggregate(), $\verb|S2|$ does not exploit data sparsity in feature matrix $\bm{H}$ since $\verb|S2|$ views $\bm{H}$ as a dense matrix.

\begin{table}[H]
\centering
\caption{Average speedup  (geometric  mean)}
\begin{adjustbox}{max width=0.48\textwidth}
\begin{tabular}{ccccc}
\toprule
\textbf{Sparsity of weight matrices} & $<50\%$ & $50\%-70\%$ & $70\%-90\%$ &  $>90\%$  \\ \midrule \midrule
\textbf{SO-S1} & $2.16\times$ & $4.36\times$ &  $10.77\times$ & $15.96\times$  \\ \midrule
\textbf{SO-S2} & $1.38\times$ & $1.64\times$ &  $2.11\times$ & $5.03\times$  \\ \bottomrule
\end{tabular}
\end{adjustbox}
\label{tab:average-speedup}
\end{table}

{
In conclusion, the proposed dynamic K2P mapping strategy leads to lower accelerator execution latency compared with the static mapping strategies. Using dynamic K2P mapping strategy, the execution latency reduces as the data sparsity increases. }




\subsection{Analysis of Compiler and Runtime System}
\label{subsec:analysis-compiler-runtime}

\begin{table}[h]
\centering
\caption{The preprocessing time of the compiler (ms)}
\begin{adjustbox}{max width=0.48\textwidth}
\begin{tabular}{c|cccccc}
\toprule
 & CI & CO & PU & FL & NE & RE \\ \midrule \midrule
GCN & 2.5E-1 & 2.2E-2  & 5.7E-1 & 2.68E0 & 1.70E0 & 5.1E1\\ 
GraphSAGE  & 2.3E-1 & 2.6E-1 & 5.9E-1 & 2.58E0 & 1.65E0 & 4.9E1 \\  
GIN &2.4E-1& 2.6E-3&5.8E-1& 2.69E0&1.71E0&5.0E1 \\
SGC &2.3E-1& 2.4E-3&6.1E-1& 2.74E0&1.73E0&5.2E1 \\
 \bottomrule
\end{tabular}
\label{tab:overhead-of-compiler}
\end{adjustbox}
\end{table}

\noindent \textbf{Overhead of the compilation/preprocessing}: Table \ref{tab:overhead-of-compiler} shows the overhead
of the compiler on the host processor (Intel Xeon 5120). The processing time includes the overheads of generating IR, data partitioning, and preprocessing of data sparsity. Compared with design automation framework \cite{liang2020deepburning} which needs to regenerate FPGA accelerator if the graph or GNN model changes, the overhead of the compiler in our design is small.

\begin{figure}[h]
     \centering
      \includegraphics[width= 8cm]{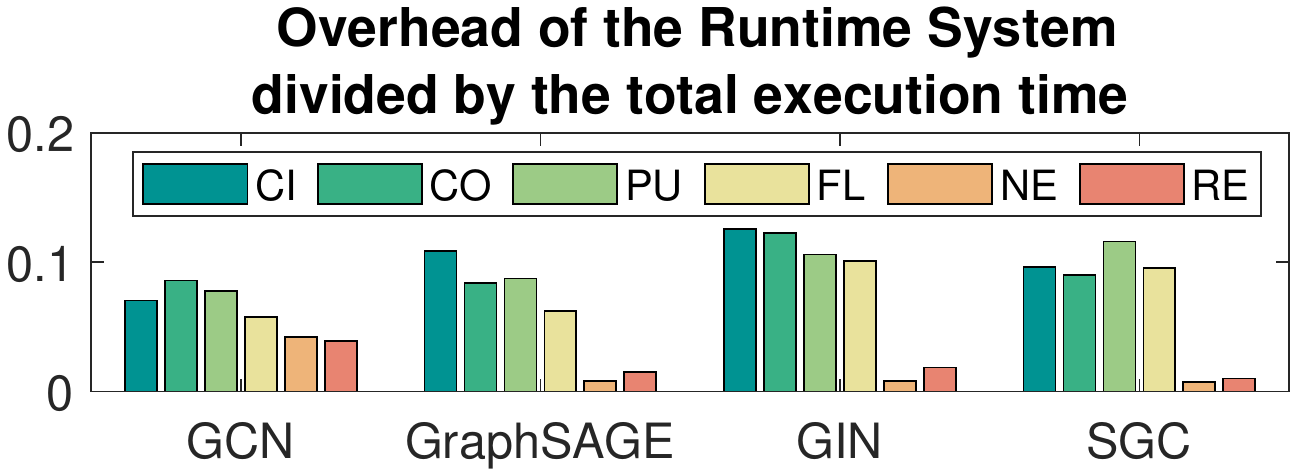}
     \caption{Overhead of runtime system on unpruned GNNs}
     
     \label{fig:overhead-of-runtime}
\end{figure}

\noindent \textbf{Overhead of the Runtime System}: We measure the overhead of the runtime system, which is the execution time of dynamic K2P mapping on the soft processor. See Figure \ref{fig:overhead-of-runtime}.  on the average, the Runtime System takes $6.8\%$ of the total execution time and is hidden by the task scheduling (Section \ref{subsec:task-scheduling}). For the pruned GNN models, as the densities of weight matrices decrease, the overhead of the Runtime System will decrease since there will be more empty data partitions skipped by the runtime system (Algorithm \ref{alg:Kernal-to-mapping-Algorithm}).

\subsection{Comparison with the State-of-the-art}
\label{subsec:comparison-with-SOTA}

\begin{figure}[h]
     \centering
      \includegraphics[width=8.5cm]{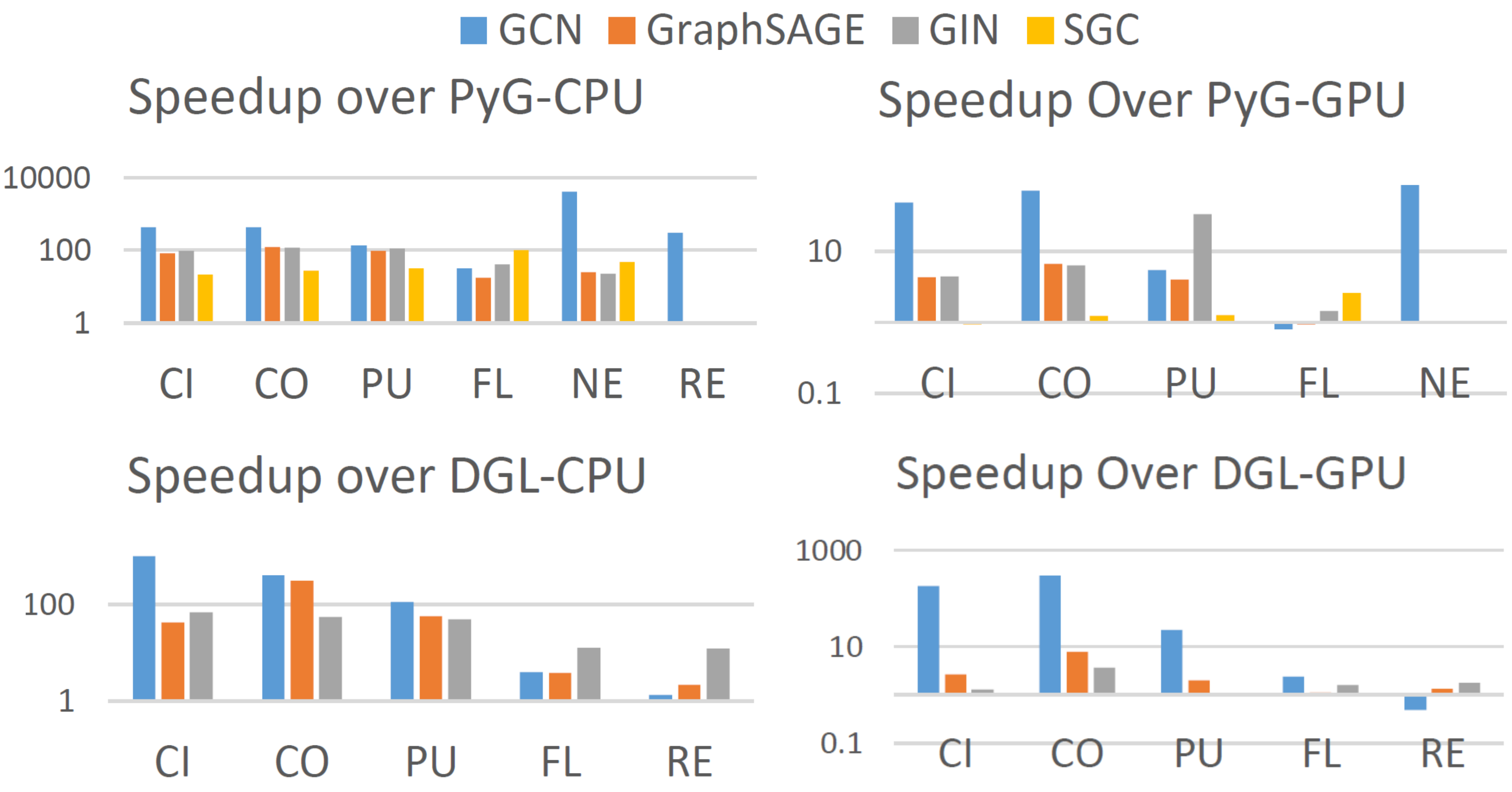}
      \caption{Speedup over the CPU and GPU platforms (Some results are not shown due to out of memory on CPU/GPU)}
     \label{fig:compare-CPU-GPU}
 \end{figure}

\noindent \textbf{Comparison with CPU/GPU}: We execute the state-of-the-art GNN frameworks -- Pytorch Geometric ({PyG, version 1.11.0}) and Deep Graph library ({DGL, version 0.8.0post2}) on CPU and GPU platforms (Table \ref{tab:platform-specifications}).  The evaluation results are shown in Figure \ref{fig:compare-CPU-GPU}. We execute the same unpruned GNN models on CPU, GPU and Dynasparse for a fair comparison. Dynasparse achieves $306\times$, $16.4\times$, $141.9\times$ and $35\times$ speedup compared with PyG-CPU, PyG-GPU, DGL-CPU and DGL-GPU, respectively. Note the CPU and GPU have $7.2\times$ and $70\times$ higher peak performance than Dynasparse. The achieved speedup is because Dynasparse can efficiently exploit the data sparsity in graph structure, vertex features and weight matrices. In contrast,  PyG and DGL on CPU and GPU only exploit the sparsity in the graph structure. {Moreover, Dynasparse exploits FPGA-specific optimizations: (1) customized data-path with Index/Data Shuffle Networks to handle the irregular memory access pattern of GNNs, (2) customized on-chip memory management for exploiting data locality, (3) dedicated hardware modules for sparsity profiling and layout/format transformation; The proposed double buffering hides their overheads, and (4) lightweight soft processor interacting with Computation Cores with extreme low latency for dynamic kernel-to-primitive mapping.}

\begin{table}[h]
\centering
\caption{Comparison of latency with the state-of-the-art GNN accelerators (using GCN model)}
\begin{adjustbox}{max width=0.48\textwidth}
\begin{threeparttable}[t]
\begin{tabular}{c|cccccc|c}
\toprule
 & CI & CO & PU & FL & NE & RE & \begin{tabular}[|c|]{@{}c@{}}  Peak Perf. \\ (TFLOPS) 
 \end{tabular}\\ \midrule \midrule
BoostGCN \cite{zhang2021boostgcn} & 1.9E-2 & 2.5E-2  & 1.6E-1 & 4.0E1 & N/A$\star$ & 1.9E2 & 1.35\\ 
HyGCN \cite{yan2020hygcn} & 2.1E-2 & 3E-1 & 6.4E1 & N/A & N/A & 2.9E2 & 4.6 \\  
\midrule
Dynasparse &7.7E-3& 4.7E-3&6.3E-2& 8.8E0&2.9E0&1.0E2 &0.512 \\
 \bottomrule
\end{tabular}
\begin{tablenotes}
\item[] $\star$ N/A: not available.
 \end{tablenotes}
\end{threeparttable}
\label{tab:comparison-sOTA}
\end{adjustbox}
\end{table}

\noindent \textbf{Comparison with GNN accelerators}: Table \ref{tab:comparison-sOTA} shows the comparison of the latency with the state-of-the-art GNN accelerators,  {  which do not require regenerating accelerator if the data sparsity changes.} All the accelerators execute the same unpruned GCN models and graph datasets. Dynasparse achieves $2.7\times$, $171\times$  speedup on the average than BoostGCN and HyGCN,  respectively. The platforms used in BoostGCN and HyGCN have $1.25\times$, $9\times$ higher peak performance than Dynasparse. The achieved speedup is because Dynasparse can efficiently exploit data sparsity in vertex features. We expect to achieve higher speedup when executing the same pruned GNN models, since \cite{zhang2021boostgcn, yan2020hygcn} do not exploit the sparsity in weights.

 {
\vspace{0.1cm}
\noindent \textbf{Discussion of preprocessing and data communication overheads}: We define the \emph{end-to-end latency} as the sum of \textbf{(1)} the overhead of compilation/preprocessing (Section \ref{subsec:analysis-compiler-runtime}), \textbf{(2)} the overhead of CPU-FPGA data movement (moving the processed input graph, processed GNN model, and optimized IR from the host memory to FPGA external memory), and \textbf{(3)} execution latency of the accelerator. With respect to end-to-end latency, Dynasparse still achieves $56.9\times$, $2.37\times$, $16.3\times$, $1.37\times$ speedup on the unpruned GNN models compared with PyG-CPU, PyG-GPU, DGL-CPU, and DGL-GPU, respectively. 
The preprocessing overhead, data movement overhead, and execution latency contribute to $43.1\%$, $27.2\%$, $27.6\%$ of the total end-to-end latency on the average. The major overhead in preprocessing is data partitioning that reorganizes the input data into data partitions. It can be reduced by multi-threading and increasing the host memory bandwidth.

Note that the CPU-FPGA data movement overhead depends on the PCIe bandwidth. The sustained PCIe bandwidth of the Alveo U250 FPGA board is around $11.2$ GB/s while the baseline GPU (Nvidia RTX3090) has PCIe bandwidth of $31.5$ GB/s. The overhead of CPU-FPGA data movement can be reduced by exploiting state-of-the-art CPU-FPGA interconnection techniques (offered by FPGA vendors), such as PCIe 5.0. Since prior GNN accelerators \cite{zhang2021boostgcn, yan2020hygcn} do not include their preprocessing overheads (data partitioning and CPU-FPGA data movement), in Table \ref{tab:comparison-sOTA}, we only compare the accelerator execution latency with \cite{zhang2021boostgcn, yan2020hygcn} for a fair comparison. 
}

%% file: 9-conclusion.tex
 \section{Conclusion and Future Work}

In this paper, we proposed a hardware-software codesign for dynamic sparsity exploitation in GNN inference.  The proposed dynamic K2P mapping reduces the inference latency by $3.73\times$ on the average compared with the static mapping strategies.  Compared with state-of-the-art CPU (GPU) implementations, Dynasparse achieves up to $56.9\times$ ($2.37\times$)  speedup in end-to-end latency. Compared with state-of-the-art FPGA implementations, Dynasparse achieves $2.7\times$ speedup in accelerator execution latency.    In the future, we plan to extend Dynasparse on heterogeneous platforms that consist of CPU, GPU and FPGA, where GPU is effective for dense primitives, FPGA is effective for sparse primitives and the CPU can execute complex control flow (e.g., dynamic K2P mapping).  

{
\section*{Acknowledgment}
This work is supported by the National Science Foundation (NSF) under grants CCF-1919289 and OAC-2209563. Equipment
and support by Xilinx are greatly appreciated.}

%% file: main.bbl
\begin{thebibliography}{10}
\providecommand{\url}[1]{#1}
\csname url@samestyle\endcsname
\providecommand{\newblock}{\relax}
\providecommand{\bibinfo}[2]{#2}
\providecommand{\BIBentrySTDinterwordspacing}{\spaceskip=0pt\relax}
\providecommand{\BIBentryALTinterwordstretchfactor}{4}
\providecommand{\BIBentryALTinterwordspacing}{\spaceskip=\fontdimen2\font plus
\BIBentryALTinterwordstretchfactor\fontdimen3\font minus
  \fontdimen4\font\relax}
\providecommand{\BIBforeignlanguage}[2]{{%
\expandafter\ifx\csname l@#1\endcsname\relax
\typeout{** WARNING: IEEEtran.bst: No hyphenation pattern has been}%
\typeout{** loaded for the language `#1'. Using the pattern for}%
\typeout{** the default language instead.}%
\else
\language=\csname l@#1\endcsname
\fi
#2}}
\providecommand{\BIBdecl}{\relax}
\BIBdecl

\bibitem{zhong2022explainable}
W.~Zhong and P.~Balaprakash, ``Explainable graph pyramid autoformer for
  long-term traffic forecasting,'' \emph{arXiv preprint arXiv:2209.13123}.

\bibitem{hewes2021graph}
J.~Hewes, ``Graph neural network for object reconstruction in liquid argon time
  projection chambers,'' in \emph{EPJ Web of Conferences}.

\bibitem{yan2020hygcn}
M.~Yan and L.~Deng, ``Hygcn: A gcn accelerator with hybrid architecture,'' in
  \emph{2020 IEEE International Symposium on High Performance Computer
  Architecture (HPCA)}.\hskip 1em plus 0.5em minus 0.4em\relax IEEE, 2020.

\bibitem{zhang2021boostgcn}
B.~Zhang, R.~Kannan, and V.~Prasanna, ``Boostgcn: A framework for optimizing
  gcn inference on fpga,'' in \emph{2021 FCCM}.\hskip 1em plus 0.5em minus
  0.4em\relax IEEE.

\bibitem{zhang2020hardware}
B.~Zhang, H.~Zeng, and V.~Prasanna, ``Hardware acceleration of large scale gcn
  inference,'' in \emph{2020 IEEE ASAP}, pp. 61--68.

\bibitem{lin2022hp}
Y.-C. Lin, B.~Zhang, and V.~Prasanna, ``Hp-gnn: generating high throughput gnn
  training implementation on cpu-fpga heterogeneous platform,'' in
  \emph{Proceedings of the 2022 ACM/SIGDA International Symposium on
  Field-Programmable Gate Arrays}, 2022, pp. 123--133.

\bibitem{meng2021dynamap}
Y.~Meng, S.~Kuppannagari, R.~Kannan, and V.~Prasanna, ``Dynamap: Dynamic
  algorithm mapping framework for low latency cnn inference,'' in \emph{The
  2021 ACM/SIGDA International Symposium on Field-Programmable Gate Arrays},
  2021, pp. 183--193.

\bibitem{zhang2023graphagile}
B.~Zhang, H.~Zeng, and V.~Prasanna, ``Graphagile: An fpga-based overlay
  accelerator for low-latency gnn inference,'' \emph{arXiv preprint
  arXiv:2302.01769}, 2023.

\bibitem{sarkar2022flowgnn}
R.~Sarkar, S.~Abi-Karam, Y.~He, L.~Sathidevi, and C.~Hao, ``Flowgnn: A dataflow
  architecture for universal graph neural network inference via multi-queue
  streaming,'' \emph{arXiv preprint arXiv:2204.13103}, 2022.

\bibitem{kipf2016semi}
T.~N. Kipf and M.~Welling, ``Semi-supervised classification with graph
  convolutional networks,'' \emph{arXiv preprint arXiv:1609.02907}, 2016.

\bibitem{hamilton2017inductive}
W.~L. Hamilton, R.~Ying, and J.~Leskovec, ``Inductive representation learning
  on large graphs,'' in \emph{Proceedings of the 31st NeurIPS}.

\bibitem{xu2018powerful}
K.~Xu, W.~Hu, J.~Leskovec, and S.~Jegelka, ``How powerful are graph neural
  networks?'' \emph{arXiv preprint arXiv:1810.00826}, 2018.

\bibitem{wu2019simplifying}
F.~Wu and A.~Souza, ``Simplifying graph convolutional networks,'' in
  \emph{International conference on machine learning}.\hskip 1em plus 0.5em
  minus 0.4em\relax PMLR, 2019.

\bibitem{pyg-dataset}
\BIBentryALTinterwordspacing
``graph datasets.'' [Online]. Available:
  \url{https://pytorch-geometric.readthedocs.io/en/latest/modules/datasets.html}
\BIBentrySTDinterwordspacing

\bibitem{rahman2022triple}
M.~Rahman, A.~Azad \emph{et~al.}, ``Triple sparsification of graph
  convolutional networks without sacrificing the accuracy.''

\bibitem{chen2021unified}
T.~Chen and Y.~Sui, ``A unified lottery ticket hypothesis for graph neural
  networks,'' in \emph{ICML}, 2021.

\bibitem{geng2020awb}
T.~Geng and A.~Li, ``Awb-gcn: A graph convolutional network accelerator with
  runtime workload rebalancing,'' in \emph{2020 IEEE/ACM MICRO}, 2020.

\bibitem{liang2020deepburning}
S.~Liang, ``Deepburning-gl: an automated framework for generating graph neural
  network accelerators,'' in \emph{2020 IEEE/ACM ICCAD}.

\bibitem{chen2015energy}
R.~Chen, S.~Siriyal, and V.~Prasanna, ``Energy and memory efficient mapping of
  bitonic sorting on fpga,'' in \emph{2015 ACM/SIGDA FPGA}.

\bibitem{lakhotia2020gpop}
K.~Lakhotia, R.~Kannan, S.~Pati, and V.~Prasanna, ``Gpop: A scalable cache-and
  memory-efficient framework for graph processing over parts,'' \emph{ACM
  Transactions on Parallel Computing (TOPC)}, 2020.

\bibitem{ref-alvelu250}
\BIBentryALTinterwordspacing
``Xilinx alveo u250.'' [Online]. Available:
  \url{https://docs.xilinx.com/r/en-US/ds962-u200-u250/FPGA-Resource-Information}
\BIBentrySTDinterwordspacing

\bibitem{Microblaze-link}
\BIBentryALTinterwordspacing
``Microblaze.'' [Online]. Available:
  \url{https://docs.xilinx.com/v/u/2021.1-English/ug984-vivado-microblaze-ref}
\BIBentrySTDinterwordspacing

\bibitem{zeng2019graphsaint}
H.~Zeng, H.~Zhou, A.~Srivastava, R.~Kannan, and V.~Prasanna, ``{GraphSAINT}:
  Graph sampling based inductive learning method,'' in \emph{International
  Conference on Learning Representations}.

\bibitem{yang2016revisiting}
Z.~Yang, W.~Cohen, and R.~Salakhudinov, ``Revisiting semi-supervised learning
  with graph embeddings,'' in \emph{International conference on machine
  learning}.\hskip 1em plus 0.5em minus 0.4em\relax PMLR, 2016, pp. 40--48.

\end{thebibliography}
